\documentclass[preprint]{aastex631}

\accepted{December 1, 2021}

\submitjournal{AJ}

\shorttitle{Challenging DIM: YZ\,LMi}
\shortauthors{Baptista \& Schlindwein}
\graphicspath{{./}{figures/}}

\begin{document}

\title{Challenging the disk instability model: \\
        I - The case of YZ\,LMi}

 \correspondingauthor{Raymundo Baptista}
 \email{raybap@gmail.com}


\author[0000-0001-5755-7000]{Raymundo Baptista}
\affiliation{Departamento de F\'{i}sica \\
  Universidade Federal de Santa Catarina \\
  Campus Trindade, Florian\'{o}polis, SC, Brazil}

\author{Wagner Schlindwein}
\affiliation{Departamento de F\'{i}sica \\
  Universidade Federal de Santa Catarina \\
  Campus Trindade, Florian\'{o}polis, SC, Brazil}



\begin{abstract}

Observations of YZ\,LMi show enhanced emission along the stream
trajectory beyond impact at disk rim during outbursts as well as when
the quiescent disk is large. We investigated whether these features can
be explained in terms of either gas stream overflow or penetration
within the frameworks of the disk-instability (DIM) and the mass-transfer
instability (MTIM) models of outbursting disks. Gas stream overflow
is not possible because the vertical scaleheight of the
stream is significantly lower than that of the outer disk and because
there is no combination of parameters which enables stream overflow on a
larger disk while preventing it on a smaller disk. Stream penetration
requires the gas stream to be denser than the outer disk regions. This
requirement cannot be met by a low-viscosity DIM disk because its density
is significantly larger than that of the gas stream over the whole range
of mass transfer rates where the thermal-viscous instability occurs.
On the other hand, the high-viscosity MTIM disk has much lower densities
which decrease with increasing radius, easily allowing for gas stream
penetration during outbursts (when mass transfer rate and stream density
increase) as well as in large quiescent disks.
The observed features are not consistent with DIM, but can be plausibly
explained by MTIM. These results suggest that the outbursts of YZ\,LMi
are the response of a high-viscosity disk to bursts of enhanced mass
transfer rate. In this case, the outburst decline timescale of (2-3)\,d
implies a viscosity parameter in the range $\alpha=3-4$.

\end{abstract}




\section{Introduction} \label{sec:intro}

Dwarf Novae are compact binaries (orbital periods $80~\mathrm{min} <
P_\mathrm{orb} \leq 8$\,h) in which a low-mass, late-type donor star feeds
hydrogen-rich gas to a companion white dwarf via an accretion disk. They
show recurrent outbursts on timescales of days-months, in which their
accretion disks brighten by factors 10-100 ($\simeq 2$-5 mag amplitude)
during a few to several days \citep{warner2003}.
The short-period dwarf novae of the SU\,UMa type ($P_\mathrm{orb}< 2$\,h)
show additional longer, brighter and more regular superoutbursts
characterized by the presence of a brightness modulation with period
slightly larger than orbital (the {\em superhump}), which is best
understood in terms of the tidal interaction between the mass donor star
and an elliptical, slowly precessing outer disk excited when the disk expands
beyond the 3:1 resonance radius \citep{whitehurst88,Hirose1990,lubow94}.
At the low end of the mass transfer rate range, the dwarf novae of the
WZ\,Sge type only show superoutbursts, with very long recurrence timescales
\citep[years to decades,][]{hellier01,warner2003}.

Outbursting AM Canum Venaticorum (AM\,CVn) systems are the ultracompact
($P_\mathrm{orb} < 65$\,min), hydrogen-deficient cousins of dwarf novae,
hosting very low-mass, at least partially degenerate donor stars
\citep[see e.g.,][] {Nelemans2005,Ramsay2007,Roelofs2010}. Most (if not
all) outbursting AM\,CVn show superoutbursts, with durations of about
9-20\,days, recurrence times from $\sim 45$ up to $\sim 450$ days
\citep{Levitanetal2011, Ramseyetal2012}, and superhumps which, in some
cases, are still present a few weeks after the end of the superoutburst
\citep[e.g.,][hereafter C11]{Copperwheat2011}.
Significant changes in recurrence time were observed at least in two
systems \citep[CR\,Boo and KL\,Dra,][]{Katoetal2001,Ramseyetal2012}.

Outbursts of dwarf novae and AM\,CVn systems may be explained in terms of
either a thermal-viscous disk-instability model \cite[DIM, e.g.,][and
references therein]{cannizzo93,lasota01} or a mass-transfer instability
model \cite[MTIM, e.g.,][]{bath,BathPringle81}. DIM predicts matter
accumulates in a cold, low-viscosity
\footnote{here we adopt the prescription of \cite{ss} for
  the accretion disk viscosity, $\nu = \alpha c_s H$, where
  $\alpha$ is the non-dimensional viscosity parameter, $c_s$ is
  the local sound speed and $H$ is the disk scaleheight.}
disk during quiescence ($\alpha_c \sim 10^{-2}$) and switches to a hot,
higher viscosity regime ($\alpha_h \sim [5-10] \alpha_c$) during outbursts,
whereas in MTIM the outburst is the response to a burst of enhanced mass
transfer rate by a disk of constant, high viscosity \cite[$\alpha\sim 1-3$,
from the decline timescale of outbursting dwarf novae, e.g.,][]
{mb83,warner2003}.
Quiescence offers good prospects for critically testing the proposed
accretion disk outburst models because it is where differences between the
predictions of both models are largest and, therefore, easier to distinguish.
For example, the densities of a quiescent DIM disk are at least two orders
of magnitude larger than those of a steady-state quiescent MTIM disk, and
continously increase as the next outburst approaches. As we will see in
Sect.~\ref{penetration}, this affects how the gas stream interacts with
the outer disk in easily distinguishable ways.

While DIM became the largely dominant model to explain these outbursts,
there are several reasons to be dissatisfied with it \citep[e.g.,][]
{Smak2000}. One of the weaknesses of DIM is its prediction that dwarf novae
increase in brightness between successive outbursts as matter piles up in
the low-viscosity quiescent disk, steadily increasing its surface density,
temperature and brightness everywhere \citep[e.g.,][]{Dubus2018},
at odds with observations \cite[e.g.,][]{hellier01,warner2003}.
Furthermore, the standard DIM cannot account for superoutbursts.
The modification introduced with the thermal-tidal instability model
\citep[proposing that superoutbursts are triggered and sustained by large
increase in tidal dissipation when the disk extends beyond the 3:1
resonance radius,][]{Osaki96} has received serious criticism
\citep{Smak2009} and has several difficulties with the observations
\citep[e.g.,][and references therein]{Hameury&Lasota2005}.
The alternative modification of the enhanced mass-transfer model
\citep[EMT, proposing that superoutbursts are triggered and sustained by
a major enhancement in mass transfer rate driven by irradiation,][]
{Osaki85,Hameury&Lasota2005} has yet to show that irradiation can indeed
explain the required mass transfer rate increase \citep{VH07,VH08}.
Without irradiation, EMT basically becomes MTIM.

On the other hand, the two strongest arguments against MTIM are based
on the (incorrect) assumption that an enhanced mass transfer stream would
necessarily stop at disk rim, leading to a significant increase in
anisotropic emission from the bright spot at outburst onset (which is
unsupported by observations) as well as preventing MTIM to trigger
inside-out outbursts (which are well documented by observations).
\citet{bem83} and later \citet{bap07} pointed out that a steady-state,
high-viscosity and low density quiescent disk enables a denser gas stream
from an enhanced mass transfer burst to penetrate the outer disk regions,
creating a bright line along the ballistic stream trajectory ahead of
the disk rim and making the trailing lune of the disk significantly
brighter than its leading lune \citep[as seen in V2051~Oph,][]{watts,bb04}.
Support for this stream penetration scenario comes from numerical
simulations of mass input from a donor star onto viscous ($\alpha \sim 1$)
accretion disks \citep[e.g.,][]{bisikalo98,makita,bisikalo05}: because
the infalling gas stream is denser than the outer disk gas, there is no
bright spot at disk rim; instead, a 'hot line' forms along the ballistic
stream trajectory extending well inwards of the outer disk regions.
Gas stream penetration at outburst onset leaves no enhanced bright spot
emission footprint and, because matter can then be deposited at the inner
disk regions, can lead to inside-out outbursts for stronger bursts.
Last but not least, there is observational evidence for an increasing list
of dwarf novae the outbursts of which are inconsistent with DIM and are
seemingly powered by bursts of enhanced mass transfer, including V2051~Oph
\citep{bb04,bap07}, V4140~Sgr \citep{Baptistaetal2016}, EX~Dra and HT~Cas
\citep{bc01,bap12}, V513~Cas and IW~And \citep{hl14}, and possibly EX~Hya
\citep{hel2000}.

\object{YZ\,LMi} (=SDSS\,J0926+3624) was the first eclipsing AM~CVn star
and is one of the shortest period eclipsing binary known \citep{Anderson2005}.
Its light curve displays deep ($\sim 2$~mag) eclipses every 28.3~min,
which lasts for $\sim 2$~min, as well as $\sim 2$~mag amplitude outbursts.
The Catalina Real-Time Transient Survey data \citep[CRTS,][]{Drake2009}
shows YZ\,LMi outbursts recurring on timescales of $\sim 100-200$~days
along the 2006-2008 seasons (C11), but on a much longer timescale along
the following 5\,years, with only one recorded outburst at the end of 2012
\citep[][hereafter SB18]{SB2018}.
Based on their stellar evolutionary calculations, \citet{Deloye2007}
predicted a mass transfer rate of $\dot{M}_2\approx (7.3\pm 1.5)\times
10^{15}\,\mathrm{g\,s^{-1}}$ (corrected for the binary parameters of C11),
in agreement with the conservative mass transfer rate of $\dot{M}_2\simeq
(9.5\pm 2.4) \times 10^{15}\,\mathrm{g\,s^{-1}}$ inferred from the observed
increase in its orbital period $\dot{P}=(3.2\pm 0.4)\times10^{-13}\,
\mathrm{s\,s^{-1}}$ \citep{Szypryt2014,SB2018}.
The 2006 quiescence light curves of C11 (collected $\simeq 20$ days after
an outburst) show superhumps and an orbital hump from a compact bright spot
at an average disk rim radius of $R_d\simeq 7.7 \times 10^9\,\mathrm{cm}
\simeq 0.48\,R_\mathrm{L1}$ (where $R_\mathrm{L1}$ is the distance from the
white dwarf to the inner Lagrangian point L1). Modelling of these light
curves with assumptions independent of the binary distance leads to
average values of white dwarf (WD) mass $M_1=(0.82\pm 0.08)\,M_\odot$,
donor star mass $M_2= (0.032\pm 0.004)\,M_\odot$ and orbital separation
$a= (0.29\pm 0.01)\,R_\odot$. C11 further inferred a WD temperature of
$T_{wd}= 17000\,K$ and a corresponding distance estimate of 460--470\,pc.
The quiescence light curves of SB18 were collected close to the end
of the 4.6\,year long period without recorded outbursts and show no
evidence of either the orbital hump produced by a bright spot at
disk rim or of superhumps.
Outburst observations of YZ\,LMi show evidence of enhanced emission
along the stream trajectory beyond impact at disk rim
\citep[C11;][]{Szypryt2014}, suggesting the occurrence of gas stream
overflow or penetration on those occasions. Similar effect is seen when
the quiescent disc is large (at $R_d \simeq 1.04 \times 10^{10}\,\mathrm{cm}
\simeq 0.65\,R_\mathrm{L1}$, SB18).

This paper investigates the viability of both the gas stream overflow and
penetration explanations for the observed enhanced stream emission, within
the frameworks of the DIM and the MTIM outburst models. Sect.~\ref{context}
reviews previous work on the stream-disk interaction and sets the scene
for the testing of the stream overflow and penetration scenarios later on,
while Sect.~\ref{models} describes the MTIM and DIM disk models used in
the comparison with the gas stream vertical scaleheight and density.
Sect.~\ref{outburst} discusses the outburst properties of YZ\,LMi, the
observational evidence for enhanced stream emission during outbursts,
and puts forward an explanation of its outbursts in terms of events of
enhanced mass transfer. In Sec.~\ref{temperature} we revise the distance
estimate and the radial temperature distribution of the accretion disk.
Sect.~\ref{overflow} investigates the feasibility of the gas stream
overflow scenario, whereas Sect.~\ref{penetration} explores the
possibility of the gas stream penetration scenario. The results are
summarized and discussed in Sect.~\ref{conclusions}.

\section{The stream-disk interaction}\label{context}

\cite{ls75,ls76} investigated the dynamics of the gas stream up to its
impact with the outer disk edge.  Because gravity changes too quickly
along the ballistic trajectory, the vertical scaleheight of the stream,
$H_\mathrm{s}$, lags behind its hydrostatic equilibrium value as the
stream moves away from the L1 point, with the largest discrepancy
occurring close to the radius of the periodic orbit with the same
tangential velocity as the stream, $R_\mathrm{circ}$, where $H_\mathrm{s}$
exceeds its hydrostatic equilibrium value by factors $\simeq 3-4$.
This led to the suggestion that, if the outer disk radius coincides with
$R_\mathrm{circ}$ and its temperature is comparable to that of the gas
stream ($\leq T_2$, the mass donor star surface temperature), stream gas
at several disk scaleheights $H_\mathrm{d}$ above midplane may be able to
overflow the disk rim and proceed towards smaller radii, even when the
midplane density of the disk edge, $\rho_\mathrm{d0}(R_\mathrm{d})$, is
orders of magnitude larger than that of the gas stream, $\rho_\mathrm{s0}$
\citep{ls76,lubow89}.
This motivated a series of studies of stream-disk interaction with
$H_\mathrm{s}/H_\mathrm{d}>1$ and $\rho_\mathrm{s0}/\rho_\mathrm{d0} \ll 1$
\citep[e.g.,][]{ArmLiv96,ArmLiv98,kunze01,Godon19}.

For example, numerical simulations for $H_\mathrm{s}/H_\mathrm{d}=2$ and
$\rho_\mathrm{s0}/\rho_\mathrm{d0}= 10^{-2}$ \citep{ArmLiv98} show that
in the orbital plane ($z=0$) the denser disk flow prevails and
continues its Keplerian orbit in an almost undisturbed fashion, whereas
at large heights from mid-plane ($z\simeq 3 H_\mathrm{d}$, where
$\rho_\mathrm{s}> \rho_\mathrm{d}$) the stream flow prevails and is able to
continue its ballistic trajectory towards smaller radii. At intermediate
heights (where $\rho_\mathrm{s} \simeq \rho_\mathrm{d}$) a double shock
front leads to deflection of disk particles towards lower radii and
of stream particles towards the azimuthal direction.
In other words, the collision of a thicker, lower density gas stream with
a thinner, higher density disk leads to cross midplane disk penetration
and high-$z$ stream overflow.

\begin{figure}
  \includegraphics[width=0.75\textwidth,angle=270]{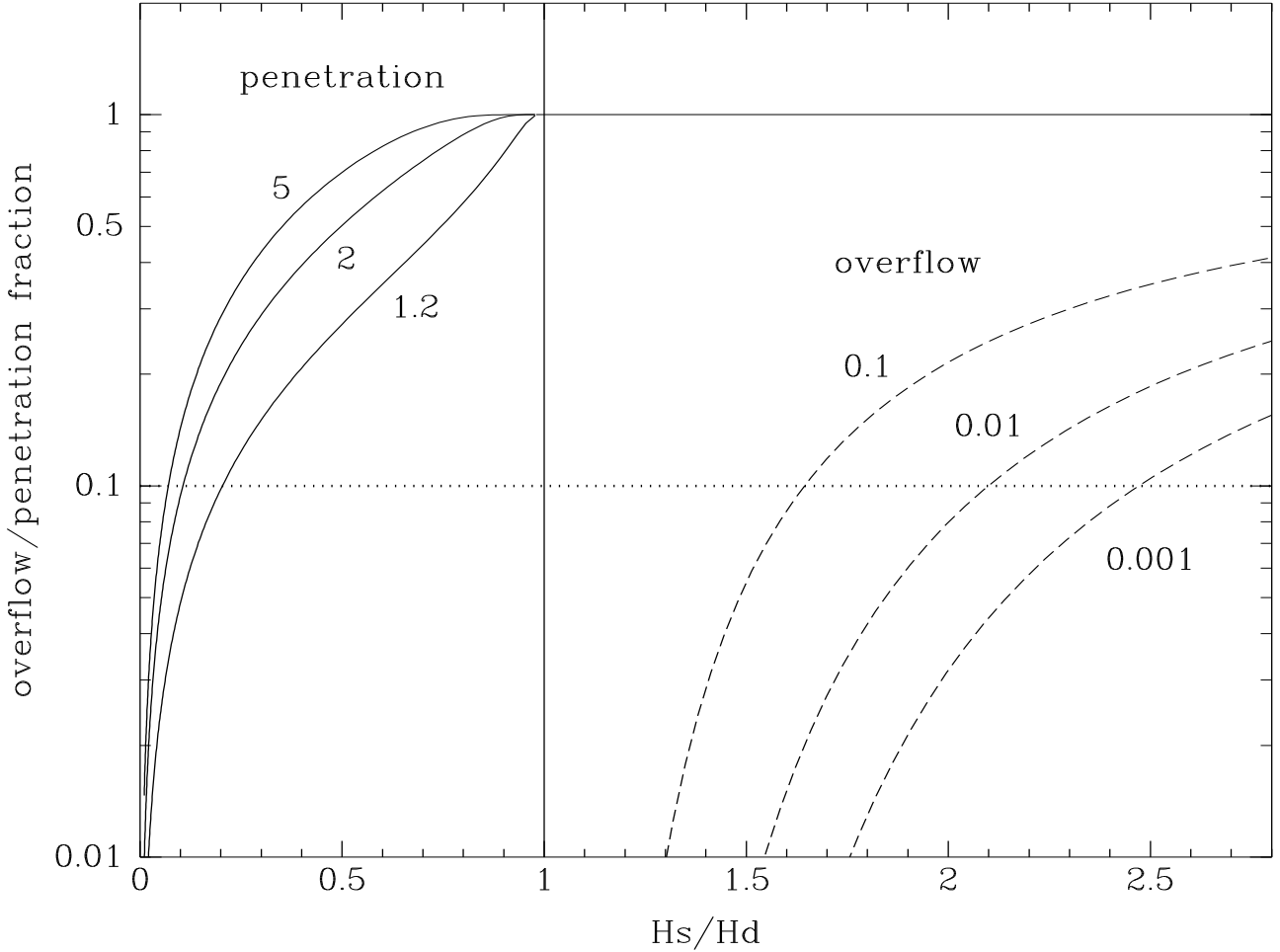}
  \caption{Fractional amount of stream overflow (dashed lines) and of
  stream penetration (solid lines) as a function of the ratio of
  scaleheights, $H_\mathrm{s}/H_\mathrm{d}$, and ratio of midplane
  densities, $\rho_\mathrm{s0}/\rho_\mathrm{d0}$ (labeled in each curve).
  \label{over-penetra}}
\end{figure}

In order to estimate the fraction of the stream mass transfer rate that
overflows the disk, $f_\mathrm{o}$, \cite{lsd86} made the very simple
assumptions of Gaussian vertical density distributions for both the disk
and the stream, and that all the mass in the stream above a critical height
$z_\mathrm{crit}$ (roughly the height at which the stream density equals
that of the disk) overflows the disk, while all stream gas below
$z_\mathrm{crit}$ is stopped by collisions with the disk \citep{hessman99},
\begin{equation}
  f_\mathrm{o} = \mathrm{erfc}
  \left[\sqrt{\frac{-\ln(\rho_\mathrm{s0}/\rho_\mathrm{d0})}
      {(H_\mathrm{s}/H_\mathrm{d})^2 - 1}}\right]  \, .
\label{eq:frac-over}
\end{equation}
Although the assumptions implied by Eq.\,(\ref{eq:frac-over}) are
arguable, \cite{ArmLiv98} found that it provides a good approximation to
the results of their simulations. Eq.\,(\ref{eq:frac-over}) is plotted
as dashed lines in Fig.\,\ref{over-penetra}.
%
%
For a $\rho_\mathrm{s0}/\rho_\mathrm{d0}\sim 10^{-2}$ density ratio, significant
stream overflow ($f_\mathrm{o}>0.1$) requires $H_\mathrm{s}/H_\mathrm{d}>2$.
No stream overflow occurs if $H_\mathrm{s}/H_\mathrm{d}<1$
\citep{ArmLiv98,hessman99}.

The stream overflow scenario of a thick stream colliding with a thin disk
\citep{ls76,lubow89} has a couple of problems.
First, measurements indicate that accretion disk radii are significantly
larger than $R_\mathrm{circ}$ \citep[by factors 2-2.5,][]{warner2003}, which
moves the stream-disk impact point out to a region where $H_\mathrm{s}$
is not particularly larger than its hydrostatic equilibrum value.
Furthermore, the midplane temperatures at the outer parts of viscous
disks ($T_c \sim \mathrm{few}\times 10^4\,K$) are typically several
times larger than the mass donor star surface temperatures ($T_2 \sim
\mathrm{few} \times 10^3\,K$), and the corresponding vertical scaleheights
are much larger than expected from hydrostatic equilibrium \citep{ho91}.
As a result, the $H_\mathrm{s}/H_\mathrm{d}$ ratios in real cases are smaller
then assumed and the condition for stream overflow is harder to occur
than envisaged. Indeed, \cite{hessman99} investigated where stream
overflow might occur in a variety of cataclysmic variables (CVs) covering
a wide range of orbital periods and disk thermal states, and found that
observationally significant ($f_\mathrm{o}> 0.1$) stream overflow should
only occur in quiescent dwarf novae with very cold outer disks.
A further misfortune makes things even less appealling.
While \cite{ls75,ls76} defined the vertical and horizontal stream
scaleheights respectively as $\chi^{-1/2}$ and $\gamma^{-1/2}$, they
tabulated values of $(2\,\pi/\chi)^{1/2}$ and $(2\,\pi/\gamma)^{1/2}$.
By taking the tabulated values in \cite{ls75,ls76} as the corresponding
vertical and horizontal scaleheights, \cite{hessman99} inadvertently
overestimated the vertical scaleheight of the gas stream by a factor
$\sqrt{2\,\pi}$ and underestimated its midplane density by a factor
$2\,\pi$. Correcting the $H_\mathrm{s}$ values of \cite{hessman99}
lowers the calculated $\log(H_\mathrm{s}/H_\mathrm{d})$ values by 0.4\,dex
(see his Fig.\,4) and brings the corresponding ratios of scaleheights
to around unity for dwarf novae and below unity in all remaining cases,
indicating that stream overflow may be rare among CVs.

The gas stream penetration scenario was first considered numerically
by \cite{bem83}, which parametrized the amount of stream stripping in
terms of the impact momentum of disk material.
They suggested that a denser gas stream penatrates the disk, leading to
a shock-heated front along the ballistic trajectory at the stream side
facing the disk flow, later confirmed by 3D numerical simulations
\citep{bisikalo98,makita,bisikalo05}.
Here we are particularly interested in the reverse situation of
the \cite{ls76} stream overflow scenario, namely, the one in which
a thin and denser gas stream collides with a thicker and lower
density accretion disk.
We may evaluate the outcome of this interaction by taking the numerical
simulations of \cite{ArmLiv98} as reference, and exchanging what we call
`stream' and `disk' flows. By analogy, we expect the denser stream flow
to prevail in the orbital plane and to continue along its ballistic
trajectory, while at large heights from midplane
($z \simeq 3\,H_\mathrm{s}$, where $\rho_\mathrm{d}>\rho_\mathrm{s}$)
the disk flow prevails and continues along its Keplerian orbits with
some vertical deflection. We therefore expect that the collision of a
thin, dense stream with a thick, lower density disk leads to cross
midplane stream penetration with high-$z$ disk overflow.

We may estimate the fraction of the stream mass transfer rate that
penetrates the disk, $f_\mathrm{p}$, following the reasoning of \cite{lsd86},
this time assuming that all mass in the stream below the critical height
$z_\mathrm{crit}$ penetrates the disk, while all stream gas above
$z_\mathrm{crit}$ is stopped by collisions with the disk. By additionally
replacing `stream' and `disk' subscripts in Eq.(\ref{eq:frac-over})
we find,
\begin{equation}
  f_\mathrm{p} = 1 - \mathrm{erfc}
  \left[\sqrt{\frac{\ln(\rho_\mathrm{s0}/\rho_\mathrm{d0})}
      {(H_\mathrm{s}/H_\mathrm{d})^{-2} - 1}}\right]  \, .
\label{eq:frac-penetra}
\end{equation}
All scepticism associated with Eq.(\ref{eq:frac-over}) also applies to
Eq.(\ref{eq:frac-penetra}). It is put forward here to illustrate the
possible outcome of the stream penetration scenario, and obviously
needs to be tested against numerical simulations which, however, are
beyond the scope of the present paper. Eq.\,(\ref{eq:frac-penetra})
is plotted as solid lines in Fig.\,\ref{over-penetra}. It predicts
(i) large penetration fractions for $H_\mathrm{s}/H_\mathrm{d} \geq 0.5$ at
reasonably small density ratios $\rho_\mathrm{s0}/\rho_\mathrm{d0} \sim 2-5$;
(ii) full stream penetration ($f_\mathrm{p}=1$) for $H_\mathrm{s}/H_\mathrm{d}
\geq 1$, provided $\rho_\mathrm{s0}/\rho_\mathrm{d0}>1$;
and (iii) that there is no gas stream penetration if
$\rho_\mathrm{s0}/\rho_\mathrm{d0}<1$.

An important difference between the stream overflow and penetration
scenarios is that the latter is a continuous process, with the outcome
of the interaction being evaluated every step along the stream trajectory
after the first impact at disk edge. How deep the stream penetrates into
the disk depends on the amount of stream stripping (i.e., how the ratio
$\rho_\mathrm{s0}/\rho_\mathrm{d0}$ changes along the trajectory) and on
the radial dependency of $H_\mathrm{s}/H_\mathrm{d}$. Given that the disk
scaleheight decreases faster than the stream scaleheight along the
ballistic trajectory (e.g., see Fig\,\ref{fig-thick}), a decrease in
$\rho_\mathrm{s0}/\rho_\mathrm{d0}$ along the stream trajectory due to
stream stripping and/or to an increase in disk density might be
compensated for by a corresponding increase in $H_\mathrm{s}/H_\mathrm{d}$,
allowing significant and deep stream penetration.

Because midplane densities decrease outwards in a viscous steady disk
while the stream midplane density increases outwards as one approaches
the L1 point (see Fig.\,\ref{fig-density}), it is useful to define the
{\em penetration radius}, $R_\mathrm{p}$, as the radius where
$\rho_\mathrm{s0}/\rho_\mathrm{d0}= 1$. If $R_\mathrm{p} > R_\mathrm{d}$
no stream penetration is possible; if $R_\mathrm{p} \leq R_\mathrm{d}$
stream penetration occurs in the radial range
$R_\mathrm{p} \leq R \leq R_\mathrm{d}$.

The collision of disk gas with the penetrating stream lead to enhanced
stream emission inwards of the outer disk, while the upper disk layers
overflowing the penetrating stream lead to a vertically-extended bulge
along the stream trajectory which may occult the inner disk regions in
high inclination systems.

\section{Accretion disk models} \label{models}

Here we describe the MTIM and DIM accretion disk models used to compute the
radial run of the disc scaleheight, $H_\mathrm{d}(R)$, and mid-plane density,
$\rho_\mathrm{d0}(R)$, to be compared against the gas stream equivalents
$H_\mathrm{s}$ and $\rho_\mathrm{s0}$ in Sects. \ref{overflow} and
\ref{penetration}.

\subsection{The MTIM disk} \label{model:mtim}

We assume the thin disk approximation and treat the vertical disc structure
as a one-dimensional version of a stellar structure. Because we are
computing quiescent disk models, we assume vertical thermal and hydrostatic
equilibria, gray atmosphere and Eddington approximations, and solve the
vertical disk structure equations \citep[e.g.,][]{Smak1984},
\begin{equation}
  dP = - \rho \Omega_k^2 z \, dz \, ,
  \label{eq:pressure}
\end{equation}
\begin{equation}
d\Sigma = 2 \rho \, dz \, ,
\end{equation}
\begin{equation}
  dF_z = \frac{3}{2} \alpha \Omega_k P \, dz \, ,
  \label{eq:flux}
\end{equation}
\begin{equation}
d\ln T = \nabla \; d\ln P \, ,
\end{equation}
\begin{equation}
  d\tau = \kappa \rho \, dz \, ,
  \label{eq:tau}
\end{equation}
where $\rho$, $P$ and $T$ are the density, pressure and temperature,
respectively, $\Sigma$ is the surface density integrated between $-z$ and
$+z$, $\Omega_k= (GM_1/R^3)^{1/2}$ is the local Keplerian frequency,
$G$ is the gravitation constant, $F_z$ is the vertical energy flux,
$\nabla$ is the temperature gradient, $\kappa$ is the frequency-averaged
opacity, and $\tau$ is the vertical optical depth.
Convection is taken into account using the mixing-length approximation
according to the prescription of \citet{paczynski}.
At the low densities of our accretion disk models ($\rho \leq 10^{-6}
\mathrm{g\, cm^{-3}}$), convection is usually superadiabatic and $\nabla$
tends to the radiative gradient,
\begin{equation}
\nabla_\mathrm{rad} = \frac{3\kappa P F_z}{16 \sigma T^4 \Omega_k^2 z} \, ,
\end{equation}
particularly in the upper disk layers.

The gray atmosphere approximation assumes that temperature varies with
optical depth as $T(\tau)^4= T_\mathrm{p}^4(1/2 + 3\tau/4)$, where the
photospheric temperature is defined as $T_\mathrm{p}= T(\tau=2/3)$.
Rosseland mean opacities, $\kappa_R$, are adequate to describe the
integrated energy flux in deep disk layers close to mid-plane,
while Planck mean opacities, $\kappa_P$, are adequate to describe the
integrated thermal emission at the atmospheric disk layers. We combine
these two opacities in a similar way to \citet{hameury98},
\begin{equation}
\kappa = \frac{\tau_e}{1+\tau_e}\kappa_R + \frac{1}{1+\tau_e}\kappa_P \, ,
\end{equation}
where $\tau_e= 1/2\,\Sigma\kappa_R$ is the estimated (mid-plane to
surface) disk optical depth. The use of $\tau_e$ instead of $\tau_e^2$
\citep{hameury98} provides a smoother transition between the two opacity
regimes in the vertical disk structure. At the surface, this leads to the
following boundary condition,
\begin{equation}
  F_z \equiv \sigma T_\mathrm{ef}^4 =
  \sigma T_\mathrm{p}^4 (1-e^{-2\tau_s}-0.84\tau_s^{3/2} e^{-2\tau_s}) \, ,
  \label{eq:surface}
\end{equation}
where $\sigma$ is the Stafan-Boltzmann constant, $T_\mathrm{ef}$ is the
effective temperature, the term in parenthesis is an analitical
approximation to the second exponential integral to better than 2 per
cent \citep{hameury98}, and $\tau_s$ is the integrated optical depth
from mid-plane to the surface (Eq. \ref{eq:tau}).
At the range of $T,\rho$ values covered by the accretion disk models,
degeneracy and radiation pressure effects are negligible. Therefore,
we complete the set of equations with the perfect gas equation of state,
\begin{equation}
  P = \frac{k}{m_H} \frac{\rho T}{\mu} =
  \frac{k}{m_H} \frac{\rho T (1+E)}{\mu_0} \, ,
  \label{eos}
\end{equation}
where $k$ is the Boltzmann constant, $m_H$ is the hydrogen atom mass, 
$\mu_0$ is the mean molecular weight for neutral gas, and $E$ is the
number of free electrons per atom. Rosseland and Planck mean opacities
and the number of free electrons per atom are obtained at the Los Alamos
TOPS Opacities site \footnote{https://aphysics2.lanl.gov/apps/}
\citep{tops}. Thermodynamical quantities 
are calculated following \citet{paczynski}.
The opacities are the most important ingredients of the model; their
influence largely overcome those of the choice of mixing-length parameter,
the treatment of the thermodynamical quantities and the switching on/off
of convection.
Calculations of the vertical disk structure with a radiative transfer
code \citep{sw91} shows that the gray atmosphere approximation is good
\citep{hameury98,idan99}.

The tidal effect of the mass donor star extracts angular momentum from
the accretion disk and truncates it at a radius \citep{warner2003},
\begin{equation}
  \frac{R_\mathrm{tid}}{R_\mathrm{L1}}=
  \frac{R_\mathrm{tid}}{a} \frac{a}{R_\mathrm{L1}}=
  \frac{0.60}{1+q} (1.0015 + q^{0.4056}) \, ,
\end{equation}
where $a$ is the orbital separation.
For YZ\,LMi (q=0.039, C11), $R_\mathrm{tid}= 0.73\,R_\mathrm{L1}$, and the
larger disk radius of SB18, $R_d \simeq 0.65\, R_\mathrm{L1}$, is close
enough to the tidal truncation radius that tidal dissipation may lead
to observable effects in the outer disk regions. Tidal torques induce
an additional viscous dissipation \citep{papa77,io92},
\begin{equation}
 Q^+_\mathrm{tid} = \frac{c\Omega_\mathrm{orb} \nu \Sigma}{4\pi}
 \left(\frac{R}{a}\right)^5 (\Omega_k-\Omega_\mathrm{orb}) \, ,
\end{equation}
where $\Omega_\mathrm{orb}= 2\pi/P_\mathrm{orb}$ is the orbital frequency
and $c$ is a non-dimensional constant adjusted to ensure that the
stationary model disk truncates at $R_\mathrm{tid}$. For YZ\,LMi we find
$c\Omega_\mathrm{orb}= 12\,\mathrm{s}^{-1}$. In thermal equilibrium, the
additional heat caused by tidal dissipation is compensated for by an
increase in the vertically-integrated flux $F^\prime_z = Q^+_\mathrm{sh}
+ Q^+_\mathrm{tid}$, where $Q^+_\mathrm{sh}= 9/8\, \nu \Sigma \Omega_k^2$
is the vertically-integrated shear viscous dissipation. The ratio of
tidal-to-shear viscous dissipation is then written \citep{io92},
\begin{equation}
  \frac{Q^+_\mathrm{tid}}{Q^+_\mathrm{sh}}= \frac{2c}{9\pi}
  \left(\frac{R}{a}\right)^5 \frac{\Omega_\mathrm{orb}}{\Omega_k}
  \left( 1 - \frac{\Omega_\mathrm{orb}}{\Omega_k} \right) \, . 
\end{equation}
We tested two different ways to add the tidal dissipation contribution
to Eq.(\ref{eq:flux}): (i) with a uniform distribution in height and
(ii) proportional to the pressure, as the shear viscous heat term
(i.e., concentrated towards disk mid-plane). Given that the results
are indistinguishable, we decided to adopt scheme (ii) as it is
computationally more robust. Therefore, tidal dissipation effects are
included in the vertical disk structure calculations by replacing
$T_\mathrm{ef} \rightarrow T_\mathrm{ef}(1+Q^+_\mathrm{tid}/Q^+_\mathrm{sh})^{1/4}$
in the boundary condition Eq.(\ref{eq:surface}) and $\alpha \rightarrow
\alpha (1+Q^+_\mathrm{tid}/Q^+_\mathrm{sh})$ in Eq.(\ref{eq:flux}).

The set of equations (\ref{eq:pressure}-\ref{eq:tau},\ref{eos}) is
integrated between the photosphere, $z=h_0$, and mid-plane, with the
boundary conditions $F_z(h_0)= \sigma T^4_\mathrm{ef}
(1+Q^+_\mathrm{tid}/Q^+_\mathrm{visc})$, $\tau(h_0)=0$,
$T(h_0)= (1/2)^{1/4} T_\mathrm{p}$ and $\rho(h_0)$ vanishingly small
(the exact choice has negligible influence in the results provided
$\rho$ is small; we adopt $\rho(h_0)= 10^{-11} \mathrm{g\,cm^3}$).
The photospheric height $h_0$ is found by a binary search algorithm
in the range $H(R)<h_0<10\,H(R)$, where $H(R)$ is an initial estimate
of the local disc scaleheight, until the convergence condition
$F_z(z=0)=0$ is satisfied. The initial iteration assumes $\tau_s\gg 1$
and $T_\mathrm{p}=T_\mathrm{eff}$; $T_\mathrm{p}$ of the subsequent
iterations is obtained from Eq.(\ref{eq:surface}). For the range of
$\alpha$ and $\dot{M}$ values of this paper, the accretion disk is
optically thick ($\tau_s \geq 10^2$) at all disk radii and
$T_\mathrm{p} \rightarrow T_\mathrm{ef}$.
Plots of the optical depth estimate
$\tau_\lambda=\Sigma\,\kappa_\lambda(T_c,\rho_c)$ using frequency dependent
opacities from the Los Alamos site show that $\tau_\lambda>1$ for
$R< 0.65\,R_\mathrm{L1}$ ($\tau_\lambda>>1$ for $R<0.5\,R_\mathrm{L1}$),
$\alpha\leq 4$ and $\dot{M}\geq 10^{15}\,\mathrm{g\,s^{-1}}$, confirming
that the MTIM accretion disk is optically thick both in the lines and in
the continuum over the range of $\alpha$ and $\dot{M}$ values of interest
(see Sections \ref{enhanced} and \ref{temperature}).

A vertical disk structure model is specified by a set of ($T_\mathrm{ef},
R, M_1, \alpha, c, P_\mathrm{orb}, a$) values and by the choice of the
chemical abundances. The relevant output quantities are $\rho_{d0}, T_c,
\tau_s$ and $\Sigma$. For an MTIM steady-state disk, the radial run of
these quantities is obtained by computing the vertical disk structure
for a set of $T_\mathrm{ef}(R)$ values given by \citep{acpower},
\begin{equation}
  \sigma T^4_\mathrm{ef}= \frac{3 G M_1 \dot{M}}{8\pi R^3}
  \left[ 1 - \left( \frac{R_1}{R} \right)^{1/2} \right] \, ,
\end{equation}
where $\dot{M}$ is the disk mass accretion rate, and the disk vertical
scaleheight is derived from,
\begin{equation}
  H_\mathrm{d}(R) = \frac{R\,c_\mathrm{s}(R)}{v_k(R)} =
  \left[ \frac{k}{GM_1m_H} \frac{T_c(R)}{\mu(R)} \right]^{1/2} R^{3/2} \, ,
 \label{eq:scaleheight}
\end{equation}
where $c_\mathrm{s}(R)$ and $v_k(R)$ are, respectively, the local sound and
Keplerian velocities, and $\mu(R)= \mu_0/[1+E(T_c(R),\rho_{d0}(R))]$ is
the mean molecular weight.
For the hydrogen-deficient disk of YZ\,LMi we assume solar metal
abundances and adopt $Y=0.98, Z=0.02, \mu_0=4.06$. Larger metalicities
increase $\kappa$, $\tau_s$, $T_c$ and $H_\mathrm{d}$ (making it harder
for gas stream overflow to occur), while decreasing $\Sigma$ and
$\rho_\mathrm{d0}$ (making it easier for gas stream penetration to occur).

\subsection{The DIM disk} \label{model:dim}

We use the DIM disk models of \cite{hameury98} and \cite{Kotkoetal2012}.
They obtain the radial structure of the disk by solving the equations
of mass, angular momentum and energy conservation, and derive the disk
local vertical structure in a way similar to that described in
Sect.\,\ref{model:mtim}. At each radius, sets of $T_\mathrm{ef},\Sigma$
values satisfying thermal equilibrium are calculated, leading to the
well-known S-shaped $T-\Sigma$ curve. The lower and upper branches of
this curve respectively represent stable cool and hot disk equilibrium
solutions, the slopes of which are well described by $T_\mathrm{ef}
\propto \Sigma^{5/14}$ \citep{Kotkoetal2012}.

In the DIM framework, mass progressively accumulates in a low-viscosity
($\alpha_c \sim 10^{-2}$) quiescent disk until the thermal-viscous
instability is triggered at a given radius, giving rise to the next
outburst. The radial run of the surface density of a quiescent DIM disk
is therefore bound to the range $\Sigma^+(R) < \Sigma(R) < \Sigma^-(R)$,
where $\Sigma^+(R)$ is the (low) critical surface density right at the
end of an outburst and $\Sigma^-(R)$ is the (high) critical surface
density at which the next outburst might be triggered. Expressions for
$\Sigma^+(R)$, $\Sigma^-(R)$ and corresponding temperatures
$T^-_\mathrm{ef}(R)$, $T^-_c$ for a He disk with solar metal abundance
($Y=0.98,Z=0.02$) are given by \citet{Kotkoetal2012},
\begin{equation}
  \Sigma^+(R)=
  380\; \alpha_{0.1}^{-0.78} m_1^{-0.35} R_{10}^{1.06}\,\,\mathrm{g\,cm^{-2}}
  \label{eq-sigmin}
\end{equation}
\begin{equation}
  \Sigma^-(R)=
  612\; \alpha_{0.1}^{-0.82} m_1^{-0.37} R_{10}^{1.10}\,\,\mathrm{g\,cm^{-2}}
  \label{eq-sigmax}
\end{equation}
\begin{equation}
  T^-_\mathrm{ef}(R)= 8690\; m_1^{-0.03} R_{10}^{-0.09}\, K
  \label{eq-temax}
\end{equation}
\begin{equation}
  T^-_c= 23600\; \alpha_{0.1}^{-0.14}\, K \, ,
  \label{eq-tc}
\end{equation}
where $\alpha_{0.1}= \alpha/0.1$, $m_1= M_1/M_\odot$ and
$R_{10}= R/10^{10}\mathrm{cm}$.
We use Eqs.\,(\ref{eq:scaleheight}) and (\ref{eq-tc}) to compute the disk
vertical scaleheight and combine it with Eqs.\,(\ref{eq-sigmin}) and
(\ref{eq-sigmax}) to derive lower and upper limit expressions for the disk
midplane density,
\begin{equation}
  \rho^+(R)= \frac{\Sigma^+(R)}{\sqrt{2\pi}\,H_\mathrm{d}(R)}
  \,\,\,\,\, , \,\,\,\,\,
  \rho^-(R)= \frac{\Sigma^-(R)}{\sqrt{2\pi}\,H_\mathrm{d}(R)}
  \,\, .
  \label{eq-rho-dim}
\end{equation}
Since $\mu$ depends on $\rho_0$, this is done iteratively.
Given that the densities steadily increase from $\rho^+$ to $\rho^-$
along quiescence, and based on our estimate that the observations of SB18
occurred $\Delta t\geq 100$ days after the previous outburst (or $\gamma=
\Delta t/T_r \geq 0.27$, see Sect.~\ref{properties}), a more restrictive
estimate for the DIM disk midplane densities at that epoch can be derived
from,
\begin{equation}
  \rho_\mathrm{med}(R) =
  \rho^+(R) + \gamma \left[ \rho^-(R) - \rho^+(R) \right] \,\, .
  \label{eq-rhomed}
\end{equation}
According to DIM, the hydrogen-deficient YZ\,LMi accretion disk will be
unstable and prone to outbursts if the mass transfer rate is in the range
$\dot{M}_2 \simeq 3 \times 10^{13} - 1.3 \times 10^{17}\,\mathrm{g\,s^{-1}}$
\citep{Kotkoetal2012}. DIM also predicts that, if $\dot{M}_2$ is large
(say, $> 5 \times 10^{15}\,\mathrm{g\,s^{-1}}$), matter accumulates in
the outer disk regions (leading to an outside-in outburst) and,
therefore, Eq.(\ref{eq-rhomed}) provides a firm lower limit to
$\rho_\mathrm{d0}(R_\mathrm{d})$. On the other hand, if $\dot{M}_2$ is
low there is enough time for the gas to diffuse inwards during
quiescence and $\rho_\mathrm{d0}(R_\mathrm{d})$ may be lower than
predicted by Eq.(\ref{eq-rhomed}). In this case,
$\rho^+(R_\mathrm{d})< \rho_\mathrm{d0}(R_\mathrm{d}) \la
\rho_\mathrm{med}(R_\mathrm{d})$.

We combine Eqs.\,(\ref{eq-sigmax})-(\ref{eq-temax}) with the slope
$T_\mathrm{ef}\propto \Sigma^{5/14}$ to derive $T_\mathrm{ef}$ for any
$\Sigma$ value in the range of interest. We gauge the influence of
tidal dissipation effects by replacing $T_\mathrm{ef} \rightarrow
T_\mathrm{ef}(1+Q^+_\mathrm{tid}/Q^+_\mathrm{sh})^{1/4}$.
The DIM disks are optical thick at all radii, with $\tau \geq 10^5$.

\section{Outburst properties of YZ\,LMi}\label{outburst}

Here we analyze the observational data on YZ\,LMi in order to estimate its
outburst length, $\Delta t$, decline timescale, $t_\mathrm{dec}$, recurrence
timescale, $T_r$, and to gauge the set of parameters within the DIM and
MTIM frameworks that describe its outburst behaviour. We also provide an
interpretation for the observed asymmetries in the eclipse shape of its
outburst light curves.

\subsection{Outburst length, decline and recurrence times} \label{properties}

The CRTS historical curve of YZ\,LMi is shown both in C11 and SB18.
The median (unfiltered) magnitude of this data set is $m= 19.35\pm 0.25$\,mag.
We defined as 'outburst' the measurements more than 3-$\sigma$ brighter
than the median magnitude, or $m(\mathrm{out})< 18.6$\,mag, and we grouped
the data by computing an average magnitude and its standard deviation per
observing night (there are up to 4 measurements per night). We further
separated these observations in two epochs, one comprising the 2006-2008
seasons where YZ\,LMi presented frequent outbursts (hereafter EP1),
and the other comprising the 2009-2013 seasons during which only one
outburst was recorded (EP2). EP1 comprises 40 groups of observations,
in 6 of which YZ\,LMi was caught in outburst. This points to an outburst
duty cycle of 15 per cent. EP2 comprises 46 groups of observations.
For a constant outburst length and recurrence timescale, one should
expect to find YZ\,LMi in outburst $\simeq 7$\,times. However, the object
was recorded in outburst only once, indicating that the recurrence time
at EP2 increased by a factor $\simeq 7$.

It is worth noting that the incomplete sampling of the CRTS data on YZ\,LMi
  \footnote{p.ex., the 2009 March outburst of C11 was not recorded by CRTS.}
does not affect the estimate of its outburst duty cycle, since the data
sampling is random (with respect to outburst state) and consistent
throughout the EP1 and EP2 epochs.

C11 observed YZ\,LMi for two consecutive nights on the decline from an
outburst on 2009 March, whereas \citet{Szypryt2014} observed YZ\,LMi over
a 4 days period along the decline of the only recorded outburst of EP2
(2012 December). Based on the eclipse shape of their light curves,
YZ\,LMi was back to quiescence on their last observing night (2012
December 11). YZ\,LMi was reported in outburst the day before the C11
observations started and, based on the measured outburst decline rate,
that outburst must have lasted until the day after their observations.
From these observations, we infer an outburst decline timescale of
$t_\mathrm{dec}\simeq (2-3)$\,days and an initial lower limit of
$\Delta t>4$\,days for the outburst length. This implies an initial lower
limit of $T_r(\mathrm{EP1})= \Delta t/0.15 > 27$\,days for the EP1 recurrence
time, indicating that the two final outburst measurements of EP1 (only 8
days apart, at Modified Julian Dates MJD 54237 and 54245), must correspond
to the same outburst. Therefore, we may improve our estimates to $\Delta t
\geq 8$\,days, $T_r(\mathrm{EP1}) \geq 53$\,days and $T_r(\mathrm{EP2})= 7\,
T_r(\mathrm{EP1}) \geq 370$\,days. At this large length these are probably
superoutbursts. Hence, YZ\,LMi seems a typical outbursting AM\,CVn star,
showing superoutbursts of length $\Delta t \geq 8$\,days and recurrence
times between 50 and $\geq 370$\,days.

In the DIM framework, the way of matching an increase in recurrence
time by a factor 7 for a given object (with fixed $M_1$, $\alpha_c$,
$\alpha_h$ and inner disk radius) is by reducing the mass transfer rate
by roughly the same amount. Hence, the observations suggest that
$\dot{M}_2(\mathrm{EP1}) \sim 7\,\dot{M}_2(\mathrm{EP2})$.
In order to match an outburst duty cycle of 15 per cent, DIM requires a
ratio $\alpha_h/\alpha_c \simeq 5$. Assuming $\alpha_h=0.1$, this leads to
$\alpha_c\simeq 0.02$. Furthermore, from the above lower limit on
$T_r(\mathrm{EP2})$, we infer that the only observed EP2 outburst started
at most on MJD 56265 and its previous outburst ended at most on MJD 55903,
a time interval of $\Delta t_\mathrm{rec} \geq 100$~days before the
observations of Paper\,I (median date of MJD 56007). Thus, the observations
of Paper\,I occurred after a fraction $\gamma = \Delta t_\mathrm{rec}/T_r
\geq 0.27$ of the YZ\,LMi outburst cycle length had elapsed. These
parameters will be used to characterize the interaction of the gas
stream and a DIM disk in Sect.~\ref{penetration}.

\subsection{Evidence for enhanced stream emission during outburst}
\label{enhanced}

The 2009 March 30 outburst observations (C11) show the asymmetric eclipse
of an extended source responsible for $\simeq 50$ per cent of the eclipsed
light when the object was $\sim 3.5$ times brighter than in quiescence.
The asymmetry had mostly vanished on the following night,
when the object dimmed to about 2 times its quiescence brightness.
C11 suggested it could arise from an uneven distribution of flux over the
disk surface or from enhanced bright spot emission and pointed out that, in
this latter case, it would imply enhanced mass transfer during the outburst.
The other remarkable feature of the light curve at that night is the
presence of a broad dip centred at phase $\simeq -0.25$, indicating the
obscuration of the white dwarf and inner disk by a vertically-extended
region of the accretion disk. C11 noted that the phasing of this dip
excludes the white dwarf, the donor star and a bright spot at disk rim as
possible occulting sources, and suggested a warped disk as its possible
cause. From the lack of evidence for the dip in the data taken on the
following night (probably the last day of the outburst), C11 concluded
that this was a very short lived feature.

We offer a different interpretation for these features.
We note that the eclipse center of the asymmetric light source is
displaced towards earlier phases and its ingress/egress features last much
longer than those of the quiescent bright spot, indicating that this
source is extended in radius and closer to disk center.
Furthermore, similar asymmetric eclipse and broad dip are also seen in the
2012 December 8 and 10 \footnote{with reduced strenght on the latter night.}
outburst light curves of \citet{Szypryt2014}, indicating these are common
and lasting features of the YZ\,LMi outbursts (unfortunately their light
curves are plotted in units of normalized flux and, thus, it is not
possible to gauge the increase in flux of these outburst light curves
with respect to their quiescent data).
We argue that the cause of the eclipse asymmetry and the dip are the same:
a radially and vertically-extended bright stream inwards of the disk rim,
responsible for a considerable fraction of the outburst light and for
obscuration of the inner disk regions over a wide range of azimuths.
Accordingly, the phasing of the dip and the eclipse asymmetry are
the same at different outbursts as well as when the effect is seen in
quiescence, in good agreement with the observations. On the other hand,
alternative explanations involving warped or tilted disks face severe
problems: (i) A warped/tilted disk would precess, leading to double wave
out-of-eclipse modulation as well as a noticeable asymmetry in eclipse
shape that would change their phase from one night to the other -- in
clear disagreement with the observations \cite[C11;][]{Szypryt2014,SB2018};
(ii) the interaction of the gas stream with a warped/tilted disk
would not lead to radially extended emission but to a bright spot inwards
of the disk rim, the radial position of which would move in and out with
disk precession phase and which would disappear beneath the accretion
disk for half of the precession cycle -- again, in clear disagreement
with the observations \cite[C11;][]{Szypryt2014,SB2018};
(iii) for an accretion disk around a $\simeq 0.82\,M_\odot$ white dwarf,
radiation-driven warping occurs only for radii $\ga 10^{14}\,\mathrm{cm}$,
several orders of magnitude larger than the size of the YZ\,LMi binary
\citep{Pringle96}.

Hence, the observations of C11 and \citet{Szypryt2014} both show
significant gas stream emission inwards of the disk rim during outbursts.
If this is taken as evidence of enhanced mass transfer during outbursts,
this suggests that the outbursts of YZ\,LMi are driven and sustained
by episodes of enhanced mass transfer, and that the system is back to
quiescence about a couple of days after the enhanced mass transfer event
is over.

In the MTIM framework, the outburst decline time is the viscous timescale
it takes for the disk to dump its excess mass onto the white dwarf
\citep{acpower,warner2003},
\begin{equation}
  t_\mathrm{dec} = t_\mathrm{visc} = \frac{R_\mathrm{d}}{v_\mathrm{R}} \sim
  \frac{R_\mathrm{d}^2}{\nu} \sim \frac{R_\mathrm{d}^2}{\alpha c_s H} \, ,
\label{eq:alpha}
\end{equation}
where $v_\mathrm{R}$ is the viscous, radial drift velocity.
In order to estimate the viscosity parameter $\alpha$ from the observed
$t_\mathrm{dec}$ range of values, we performed numerical simulations of the
response of a viscous accretion disk to a burst of enhanced mass transfer
\citep{tese} with an algorithm similar to that of \cite{io92}. Adopting
$M_1= 0.82\,M_\odot$, we need $\dot{M}_2(\mathrm{outburst})\simeq 7\,
\dot{M}_2(\mathrm{quiescence})$ in order to reproduce the observed
fractional flux amplitude of $\Delta f/f\simeq 3.5$ in the optical, and
viscosity parameters in the range $\alpha= 3-4$ in order to match the
$t_\mathrm{visc}= (2-3)$\,days requirement. This $\alpha$ range will be
used to characterize the interaction of the gas stream and an MTIM disk
in Sect.~\ref{penetration}. The inferred $\alpha$ values for YZ\,LMi are
larger than those estimated for hydrogen-rich dwarf novae accretion disks
\citep[$\alpha\simeq 1-3$,][]{mb83}, in line with the fact that $c_s$
values in a hydrogen-deficient disk are lower than those of their
hydrogen-rich counterparts by a factor $\sim 1.5$.

\section{Revised distance and radial temperature distribution}
\label{temperature}

\cite{Copperwheat2011} associated the partially eclipsed central source
to a fully visible WD at disk center (i.e., assumed an optically thin
disk) and modeled its observed UV-optical fluxes with DB synthetic spectra
\citep{boris95} to find a best-fit temperature of $T_\mathrm{wd}= 17000$\,K
and a corresponding distance estimate of 460-470\,pc to the binary.
SB18 subtracted the contribution of a 17000\,K fully visible WD from the
light curve to isolate the emission from the accretion disk in their
eclipse mapping, and assumed a distance of 460-470~pc to derive the disk
radial temperature distribution.

Here we use the trigonometric parallax $\pi= 1.227 \pm 0.208$\,mas of the
antecipated access, GAIA Data Release 3 \citep{gaia16,gaia18,lindegren}
to find a revised distance estimate to YZ\,LMi of $d= 815\pm 138$\,pc.
The estimated interstellar extinction of $A_{V+R}=0.036$\,mag \cite[from
the NASA/IPAC Infrared Science Archive site, using redenning estimates
of][]{sfd98} is negligible in comparison to the uncertainties affecting
the distance estimate and was not taken into account.
At this distance, the solid angle of the WD is reduced by a factor of
$(815/465)^2= 3.07$. In addition, the accretion disk models of
Sect.\,\ref{models} indicate that the quiescent disk of YZ\,LMi is
optically thick both in the DIM and the MTIM scenarios. As a consequence,
emission from the WD lower hemisphere is veiled by the inner opaque disk,
reducing its solid angle by an additional factor of $\simeq 2$.
This implies that the UV-optical fluxes from a 17000~K WD at the center
of an opaque accretion disk at a distance of 815~pc are a factor of
$\simeq 6$ lower than observed by C11. We are thus lead to the conclusion
that the $\simeq 17000$\,K central source is not the WD and, therefore,
it does not make sense to subtract its contribution from the light
curve to compute disk eclipse maps.
What C11 called the central source is an emitting region significantly
larger than the WD itself, perhaps the hot and opaque inner disk.

Given that we no longer know what the WD temperature and contribution
to the optical light curve are, can we constrain these quantities?
Stellar evolutionary models for AM\,CVn stars predict a WD
of $M_v\simeq 11$~mag at $P_\mathrm{orb}= 28.3$~min \citep{bildsten}.
Scaling to the 815\,pc distance of YZ\,LMi and taking into account that
half of the WD emission is veiled by the opaque disk, this translates
into a WD apparent visual magnitude of $m_v\simeq 21.3$~mag, two magnitudes
fainter than the average quiescence brightness level of YZ\,LMi,
$V\simeq 19.3$~mag. Hence, the WD is expected to be a minor contributor
and should account for about 16 per cent of the quiescent optical light.
This is in line with the lack of evidence of broad He\,I absorption wings
from a BD white dwarf in the optical spectrum of YZ\,LMi \citep{Anderson2005}.
The 16 per cent fractional contribution corresponds to a flux of $8.3\,
\mu\mathrm{Jy}$ in the V+R passband of SB18 and to a WD temperature
of 22500\,K, in good agreement with the range $T_\mathrm{wd} \simeq
(21300-23800)$\,K predicted for YZ\,LMi by stellar evolutionary
calculations of \cite{Deloye2007}.
We take these values as upper limits to the WD temperature and flux
contribution to the SB18 light curve.
Hotter WDs (with larger fractional contributions) fall into the nonradial
pulsation instability strip for isolated DBs \citep{beau99,bildsten}
and should show 200-1000\,s pulsations for which there is currently no
observational support (C11; SB18).
We then repeated the procedure of SB18 and used the YZ\,LMi eclipse
geometry to compute the light curve of a limb-darkened, half-visible
22500\,K DB white dwarf and subtracted the result from the SB18 average
light curve to separate the contribution of the remaining light sources.

\begin{figure}
  \includegraphics[width=0.75\textwidth,angle=270]{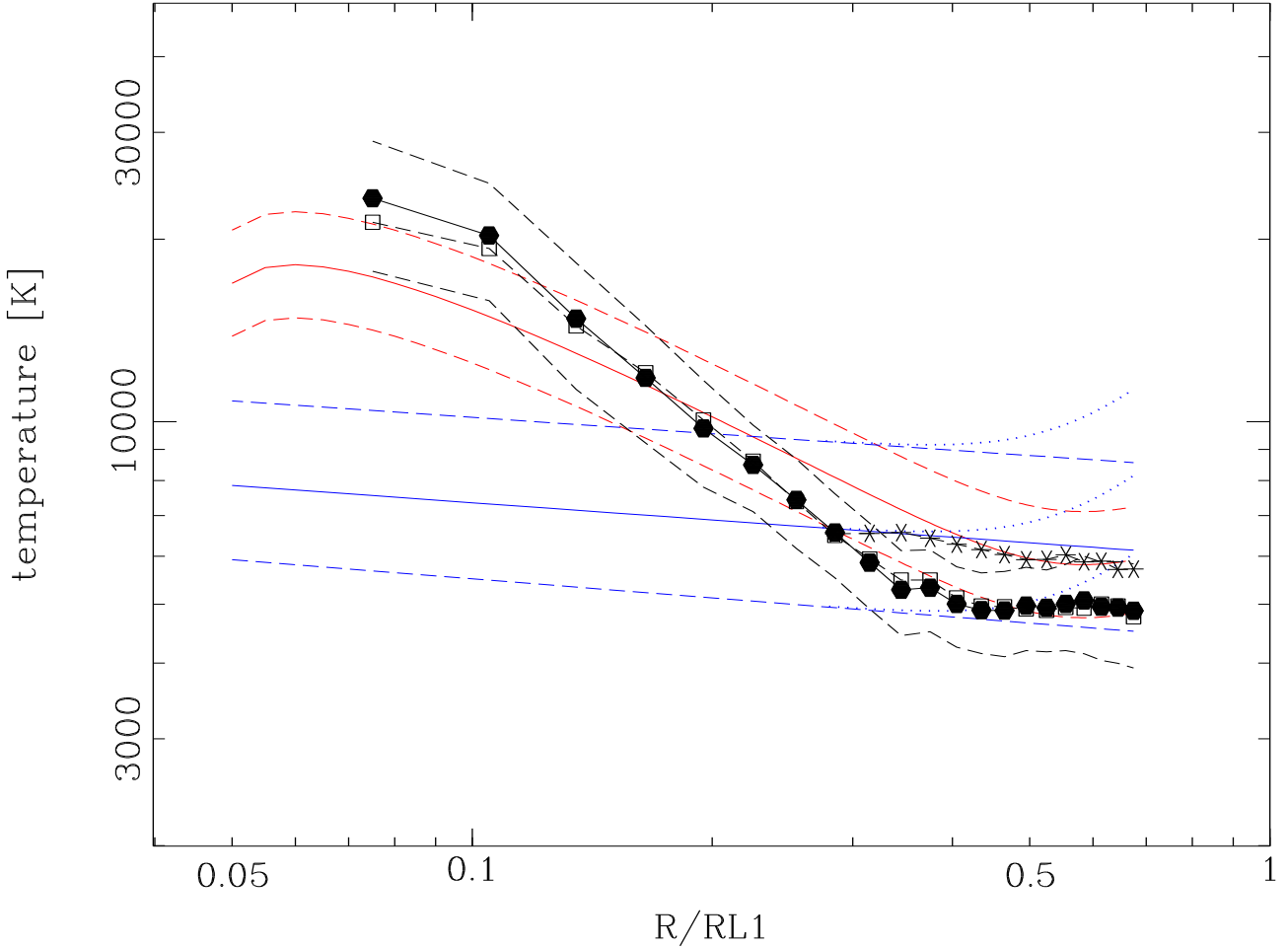}
  \caption{Azimuthally averaged radial brightness temperature distributions
  of the YZ\,LMi quiescent accretion disk without (filled circles) and with
  (open squares) subtraction of a half-visible 22500\,K WD contribution,
  for radial bins of width $0.03\,R_\mathrm{L1}$ and $d= 815\pm 138$\,pc.
  Dashed black curves show the 1-$\sigma$ limits on the distribution
  without WD subtraction. Asterisks depict temperatures along the
  ballistic stream trajectory. Red lines show $T_\mathrm{ef}(R)$ curves
  of steady-state MTIM disk models including tidal dissipation for
  $\log\dot{M}(\mathrm{g\,s^{-1}})=15.45$ (solid), 15.1 and 15.8 (dashed).
  Blue lines show $T_\mathrm{ef}(R)$ curves of DIM disk models for
  $\Sigma_\mathrm{med}$ (solid), $\Sigma^+$ and $\Sigma^-$ (dashed);
  corresponding $T_\mathrm{ef}(R)$ curves for DIM models including tidal
  dissipation are shown as blue dotted lines.
  \label{fig-trad}}
\end{figure}

Fig.\,\ref{fig-trad} shows revised radial brightness temperature
distributions of eclipse maps obtained from the average light curve of
SB18 without (filled circles) and with (open squares) subtraction of a
half-visible 22500\,K WD contribution, assuming $d= 815\pm 138$\,pc.
Error bars were derived from Monte Carlo simulations with the light curves
and take into account both the uncertainties in the distance and the
scatter of the brightness temperatures in each radial bin (SB18). Because
of the grazing eclipse, there is no information in the eclipse shape about
the brightness distribution of most of the disk hemisphere farther away
from the L1 point. Therefore, the temperature distributions were computed
from the eclipsed disk regions (mostly the disk hemisphere closest to L1).
%
%
Asterisks depict brightness temperatures for sections of azimuthal width
$30\degr$ along the ballistic stream trajectory and illustrate the
enhanced gas stream emission (SB18; see their Fig.\,4). As a consequence
of the increased distance, temperatures are sistematically larger than
those found by SB18; they range from $T_b\simeq 23000$\,K in the inner
disk to $T_b\simeq 5000$\,K in the outer disk regions. Subtracting
the (uncertain) WD contribution has only a minor effect on the radial
temperature distribution: it helps to decrease the temperatures in the
innermost disk regions and to bring the slope of the observed distribution
closer to the $T \propto R^{-3/4}$ law of steady opaque disks, but the
changes are negligible in comparison to the uncertainties.
We conclude that an estimate of the mass accretion rate from the $T(R)$
distribution is not sensitive to the uncertainty in the WD temperature
and its contribution to the optical light.

MTIM models (Sect.\,\ref{model:mtim}) predict that the disk is optically
thick ($\tau_s\geq 10^2$) at all radii for $\dot{M}\geq 10^{15}\,
\mathrm{g\,s^{-1}}$ and $\alpha\leq 4$, and DIM models
(Sect.\,\ref{model:dim}) also predict the disk is optically thick at all
radii in the range $\Sigma^+ < \Sigma < \Sigma^-$, indicating that the
YZ\,LMi observed brightness temperatures can be directly compared to
model effective temperatures in both cases without ambiguities.
These theoretical predictions are in line with the observations:
The SDSS spectrum of YZ\,LMi shows weak He\,I lines on top of a
strong blue continuum which is reasonably well fitted by a blackbody
\citep{Anderson2005}; the emission lines are minor contributors to
the optical flux. In addition, the UV-optical fluxes of the C11 central
source are well described by a 17000\,K WD spectrum as well as by a
blackbody of similar temperature (see Fig.11 of C11).
$T_\mathrm{ef}(R)$ curves for MTIM disk models with tidal dissipation
included are shown as red lines in Fig.\,\ref{fig-trad}.
While the observed temperature distribution deviates from the
$T\propto R^{-3/4}$ law at intermediate disk regions, it can be
reasonably well described by steady-state disk models in the range
$\log\dot{M}\mathrm{(g\,s^{-1})}=15.1-15.8$ at the 1-$\sigma$ limit.
This $\dot{M}$ range will be used to characterize the interaction of the
gas stream and an MTIM disk in Sects.\,\ref{overflow} and \ref{penetration}.
The flattening of the observed distribution in the outer disk regions
is well explained by tidal dissipation on a large, steady-state opaque
quiescent disk. Tidal dissipation is relevant only for $R > 0.45\,
R_\mathrm{L1}$ and is negligible when the disk is small.

$T_\mathrm{ef}(R)$ curves for DIM disk models are shown as blue lines in
Fig.\,\ref{fig-trad}. DIM predicts a flat radial temperature distribution
which underestimates the observed brightness temperatures in the inner
disk regions ($R<0.2\,R_\mathrm{L1}$). Subtracting the contribution of a
22500\,K WD from the light curve does not aleviate this discrepancy.
When tidal dissipation is included in the DIM model (dotted blue lines),
it results in increasing temperatures with radius, making the outer disk
even hotter than the inner disk regions -- in clear disagreement with the
observations.

We also obtained an updated value of $(1.4\pm 0.7)\,\mu\mathrm{Jy}$
for the uneclipsed light of the eclipse mapping procedure.

\section{The stream overflow scenario}\label{overflow}

We test the possibility of gas stream overflow by comparing the vertical
scaleheight of the disk rim, $H_\mathrm{d}(R_\mathrm{d})$, with that of
the gas stream, $H_\mathrm{s}(R_\mathrm{d})$. The condition for gas stream
overflow to occur is $H_\mathrm{s}(R_\mathrm{d}) > H_\mathrm{d}(R_\mathrm{d})$.

The gas stream vertical ($H_\mathrm{s}$) and horizontal ($W_\mathrm{s}$)
scaleheights can be obtained from the fitted expressions of
\cite{hessman99},
\begin{eqnarray}
  H_\mathrm{s}(R,q,a) & \simeq &
  \frac{a\epsilon}{\sqrt{2\pi}} \,\, h_1(R/a)\, h_2(q)\, ,
  \nonumber \\
  W_\mathrm{s}(R,q,a) & \simeq &
  \frac{a\epsilon}{\sqrt{2\pi}} \,\, w_1(R/a)\, w_2(q)\, ,
  \label{eq:hessman1}
\end{eqnarray}
where
\begin{eqnarray}
  h_1(R/a) & = & 0.060 + 3.17(R/a) - 2.90(R/a)^2 \, , \nonumber \\
  w_1(R/a) & = & 0.389 + 2.21(R/a) - 2.46(R/a)^2 \, , \nonumber \\
  \log w_2 & = & -0.023 - 0.067(\log q) + 0.081(\log q)^2\, ,\nonumber \\
  \log h_2 & = & +0.031(\log q) + 0.095 (\log q)^2 \, ,
  \label{eq:hessman2}
\end{eqnarray}
and
\begin{equation}
  a\epsilon = \frac{c_\mathrm{ss}\,P_\mathrm{orb}}{2\pi} =
  \frac{P_\mathrm{orb}}{2\pi} \,
  \left[ \frac{k\,T_2}{\mu_0\, m_\mathrm{H}} \right]^{1/2} \, ,
  \label{eq-sound}
\end{equation}
where $q=M_2/M_1$ is the binary mass ratio, $c_\mathrm{ss}$ is the average
isothermal sound speed at the L1 point, and $T_2$ is the effective
temperature of the mass donor star. Note that Eqs.(\ref{eq:hessman1})
take into account the $\sqrt{2\,\pi}$ correction factor reported in
Sect.\,\ref{context}, and that the expressions for $w_1$ and $w_2$ were
revised to take into account the updated $(2\pi/\gamma)^{1/2}$ values from
the erratum of \cite{ls14}. Combining Eqs.(\ref{eq:hessman1}-\ref{eq-sound})
we obtain,
\begin{equation}
  H_\mathrm{s}(R) \propto h_1(R/a) \, h_2(q) \, P_\mathrm{orb}\,
  \left(\frac{T_2}{\mu_0}\right)^{1/2} \, .
  \label{eq-thick-stream}
\end{equation}
For YZ\,LMi we adopt $q=0.039$, $a=0.29\,R_\odot$ (C11), $P_\mathrm{orb}=
1698.5\,s$ (SB18) and $\mu_0=4.06$. Uncertainties in the values of $q$,
$a$ and $P_\mathrm{orb}$ were not taken into account because they are
negligible in comparison to the uncertainties affecting $T_2$.

An upper limit to $T_2$ can be obtained from the
updated uneclipsed light of the eclipse mapping, $(1.4 \pm 0.7)\,
\mu\mathrm{Jy}$, the donor star radius $R_2= (0.046\pm 0.002)\, R_\odot$
(C11), and the GAIA distance estimate $815\pm 138$\,pc, under the
assumption that the mass donor star contributes 100 per cent of the
observed uneclipsed light in the $V+R$ passband
\footnote{If other sources contribute to the uneclipsed light
  \citep[p.ex., a vertically-extended disk wind,][]{Baptista2016}, the
  contribution of the mass donor star is lower than assumed and, at a
  fixed distance, its temperature is lower than the derived limit.}.
No interstellar extinction was included. A Monte Carlo simulation
independently varying the three input parameters of the fit leads to
$T_2= 3570 \pm 130\,K$; we adopt $T_2(\mathrm{max})=3700\,K$ as a
conservative upper limit to the mass donor star temperature. A lower
limit to $T_2$ can be obtained from the semi-empirical evolutionary
sequence of hydrogen-rich CV mass donor stars of \cite{knigge11}, under
the assumption that an evolved, hydrogen deficient star will be hotter
than a hydrogen-rich star of same mass. For $M_2= 0.032\,M_\odot$ we find
$T_2(\mathrm{min})= 1000\,K$. We further take the geometric average of
the upper and lower $T_2$ limits as representative of the allowed
temperature range, $T_2(\mathrm{mid})= 1900\,K$.

\begin{figure}
  \includegraphics[width=0.75\textwidth,angle=270]{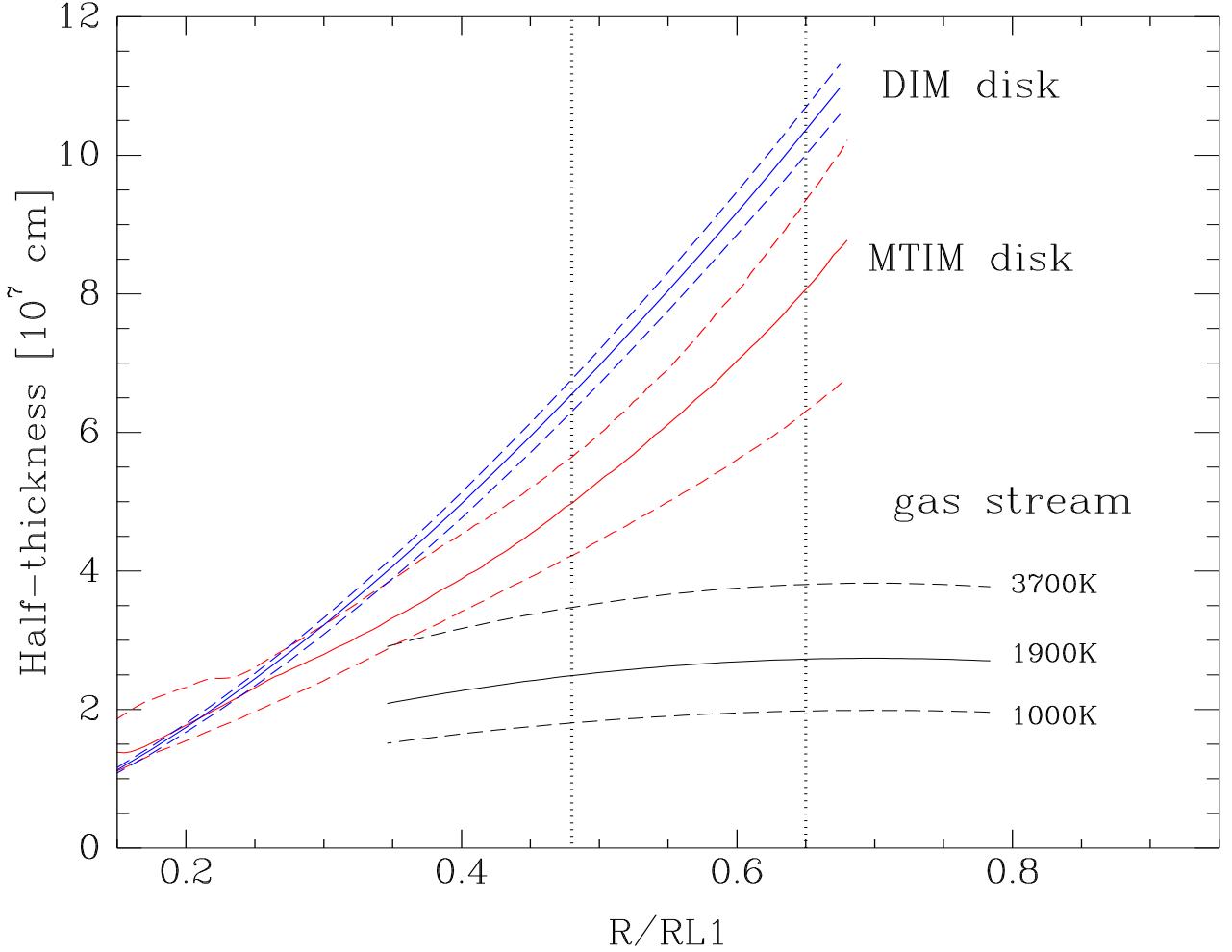}
  \caption{Comparison of the vertical scaleheights of the disk and the gas
  stream for YZ\,LMi. Radial runs of the disk scaleheight are shown for
  the DIM (blue) and MTIM (red) disk models; dashed lines show the
  corresponding 1-$\sigma$ limits. The vertical scaleheight of the gas
  stream is shown for $T_2= 1900\,K$ (solid line), $1000\,K$ and $3700\,K$
  (dashed lines). Vertical dotted lines depict the outer disk radii
  $R_d= 0.48\,R_\mathrm{L1}$ (C11) and $R_d=0.65\,R_\mathrm{L1}$ (SB18).
  \label{fig-thick}}
\end{figure}

Fig.~\ref{fig-thick} compares the vertical scaleheights of the quiescent
disk and the gas stream in YZ\,LMi. $H_\mathrm{s}(R)$ curves are shown in
black; corresponding $H_\mathrm{d}(R)$ curves for the MTIM and DIM disk models
are shown in red and blue, respectively, where we assumed $M_1=(0.82\pm
0.08)\,M_\odot$ and $\dot{M}= (1.3-6.3)\times 10^{15}\,\mathrm{g\,s^{-1}}$.
%
%
$H_\mathrm{d}(R_\mathrm{d})$ for the DIM disk is a factor of $\simeq 3$
larger than the maximum possible $H_\mathrm{s}(R_\mathrm{d})$ value at the
21-$\sigma$ level; $H_\mathrm{d}(R_\mathrm{d})$ for the MTIM disk is a factor
of 2 larger than the maximum possible $H_\mathrm{s}(R_\mathrm{d})$ value at
the 2.5-$\sigma$ level. $H_\mathrm{s}(R_\mathrm{d})$ is at least a factor
of 4 times smaller than needed for observationally significant stream
overflow to occur \citep{hessman99}.
A very hot donor star with $T_2 > 10^4\,K$ is required in order to make
$H_\mathrm{s}(R_\mathrm{d})\geq H_\mathrm{d}(R_\mathrm{d})$ and allow stream
overflow onto an MTIM disk at the lowest $\dot{M}$ value.
However, at such high temperatures the donor star would contribute
$\geq 104\,\mu\mathrm{Jy}$ to the flux in the $V+R$ passband, almost
twice the out-of-eclipse total flux and more than 5 times the
mid-eclipse flux at that passband -- in marked contradiction with the
observations. The situation is even less plausible in the DIM case.
Furthermore, because $H_\mathrm{d}(R)$ increases with radius, there is no
combination of parameters which enables stream overflow at a larger disk
radius while preventing it at a smaller radius -- as required by the
quiescence observations.

It is also worth noting that, because $H_\mathrm{d}(R_\mathrm{d})\propto
{R_\mathrm{d}}^{3/2} T_c(R_\mathrm{d})^{1/2}$, the vertical scaleheight of the
disk rim is even larger at outburst, when the disk is both larger and
hotter. On the other hand, because $H_\mathrm{s}$ does not depend on
$\dot{M}_2$, the gas stream vertical scaleheight is the same in quiescence
and in outburst, both in the DIM and the MTIM cases. Hence, excluding
the possibility of gas stream overflow in quiescence means also excluding
it in outburst.

Therefore, we are forced to discard the interpretation that the observed
enhanced gas stream emission is a consequence of stream overflow, either
within the DIM or the MTIM frameworks.

\section{The stream penetration scenario} \label{penetration}

We test the possibility of gas stream penetration by comparing the
midplane densities of the gas stream and the disk at the outer disk radius.
If $\rho_\mathrm{s0}(R_\mathrm{d}) > \rho_\mathrm{d0}(R_\mathrm{d})$ one has
$R_\mathrm{p} < R_\mathrm{d}$ and stream penetration occurs for
$R_\mathrm{p} \leq R \leq R_\mathrm{d}$.
Conversely, if $\rho_\mathrm{s0}(R_\mathrm{d}) < \rho_\mathrm{d0}(R_\mathrm{d})$
stream penetration is inhibited and the interaction of the two gas flows
leads to the formation of a shock front releasing most of the stream
kinetic energy in the form of a bright spot at disk rim.

\begin{figure}
  \includegraphics[width=0.8\textwidth,angle=0]{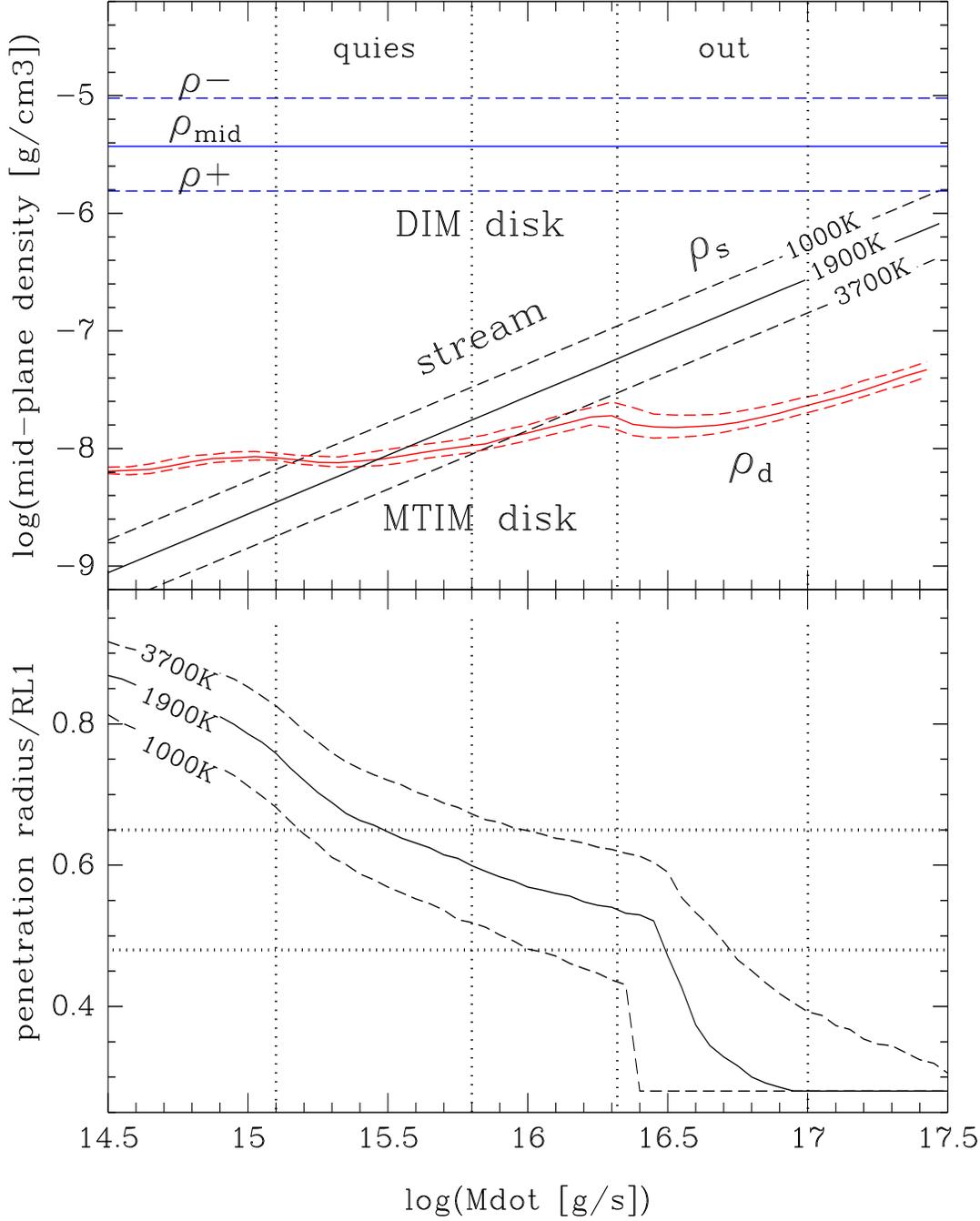}
  \caption{Top: Disk and gas stream midplane densities as a function of
    mass transfer rate for a disk radius of $R_\mathrm{d}= 0.65\,R_{L1}$.
    Red lines show the MTIM disk midplane density $\rho_\mathrm{d0}
    (R_\mathrm{d})$ for $\alpha=3.5$ (solid), 3 and 4 (dashed), while blue
    lines show the DIM disk midplane densities $\rho^+(R_\mathrm{d}),\,
    \rho^-(R_\mathrm{d})$ (dashed) and $\rho_\mathrm{med}(R_\mathrm{d})$ (solid)
    for $\alpha_c=0.02$, $\alpha_h=0.1$, and $\gamma=0.27$. Black lines
    show the gas stream midplane density $\rho_\mathrm{s0}(R_\mathrm{d})$ for
    $T_2= 1900$\,K (solid), 1000\,K and 3700\,K (dashed). Vertical dotted
    lines mark the mass transfer rate ranges $\dot{M}_2(\mathrm{quiescence})=
    (1.3-6.3)\times 10^{15}\,\mathrm{g\,s^{-1}}$ (Sect.\,\ref{temperature})
    and $\dot{M}_2(\mathrm{outburst})= 16\,\dot{M}_2(\mathrm{quiescence})=
    (2.1-10.0)\times 10^{16}\,\mathrm{g\,s^{-1}}$. Bottom: The penetration
    radius (in units of $R_\mathrm{L1}$) as a function of $\dot{M}_2$ for
    $\alpha=3.5$. Labels depict the corresponding $T_2$ value.
  \label{fig-mdot}}
\end{figure}

\begin{figure}
  \includegraphics[width=0.75\textwidth,angle=270]{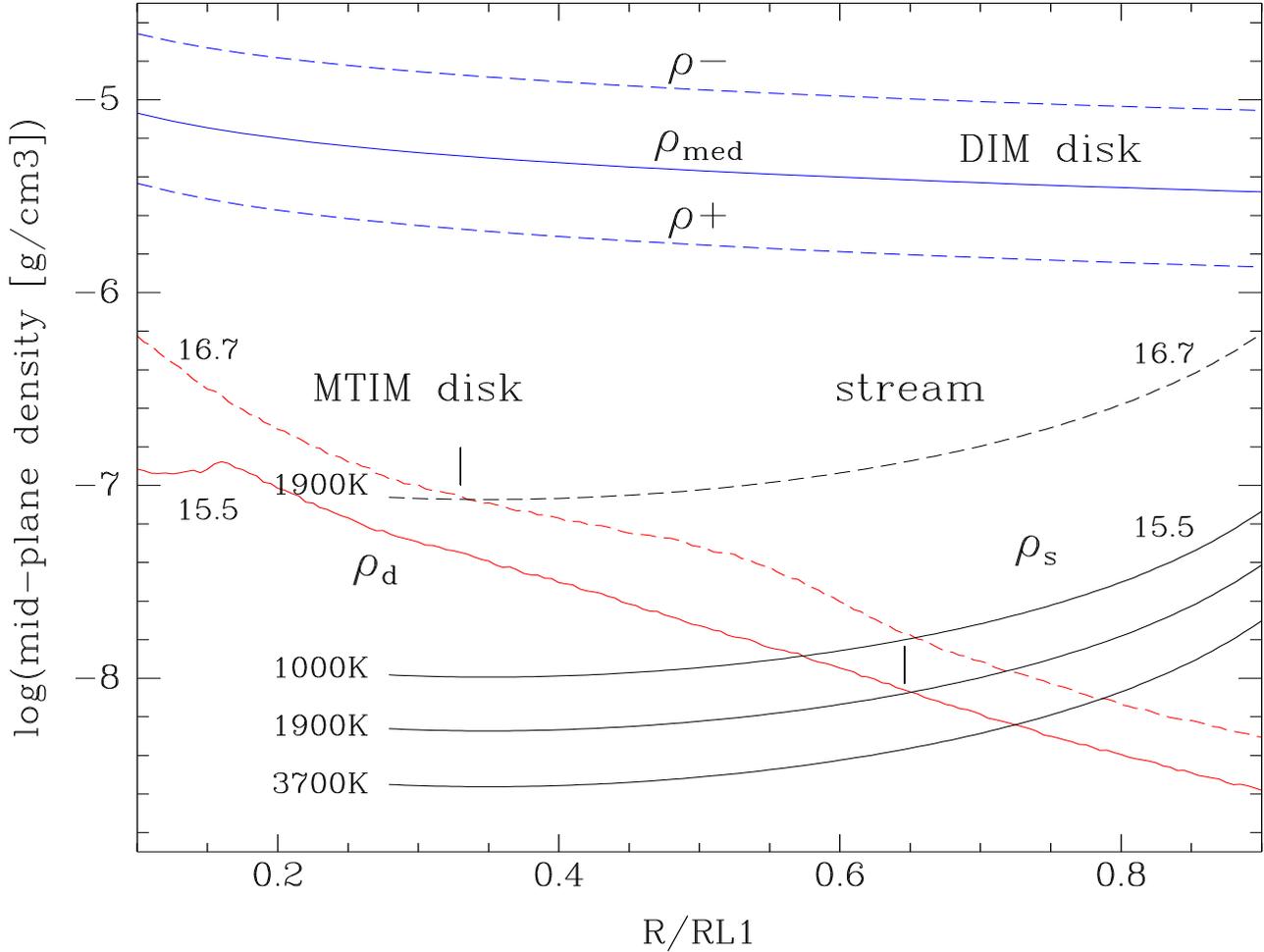}
  \caption{Disk and gas stream midplane densities as a function of disk
    radius. Red lines show the MTIM disk densities $\rho_\mathrm{d0}(R)$
    for $\alpha=3.5$, $\dot{M}_2(\mathrm{quiescence})= 3\times 10^{15}\,
    \mathrm{g\,s^{-1}}$ (solid) and $\dot{M}_2(\mathrm{outburst})= 4.8
    \times 10^{16}\,\mathrm{g\,s^{-1}}$ (dashed), while blue lines show the
    DIM disk densities $\rho^+(R),\,\rho^-(R)$ (dashed) and
    $\rho_\mathrm{med}(R)$ (solid) for $\alpha_c=0.02, \alpha_h=0.1$, and
    $\gamma=0.27$. Black lines show the gas stream midplane density
    $\rho_\mathrm{s0}(R)$ for $\dot{M}_2(\mathrm{quiescence})$ (solid) and
    $\dot{M}_2(\mathrm{outburst})$ (dashed). Labels depict the corresponding
    $T_2$ values. Vertical ticks mark the position of the penetration radius
    $R_\mathrm{p}$ at each $\dot{M}_2$ for $T_2=1900$\,K.
  \label{fig-density}}
\end{figure}

Midplane densities of MTIM disk models are plotted as a function of
$\dot{M}_2 (=\dot{M})$ in the upper panel of Fig.\,\ref{fig-mdot} for a
disk radius of $R_\mathrm{d}= 0.65\,R_\mathrm{L1}$ and as a function of
radius in Fig.\,\ref{fig-density} for mass transfer rates of
$\dot{M}_2(\mathrm{quiescence})= 3\times 10^{15}\,\mathrm{g\,s^{-1}}$ and
$\dot{M}_2(\mathrm{outburst})= 16\,\dot{M}_2(\mathrm{quiescence})= 4.8
\times 10^{16}\,\mathrm{g\,s^{-1}}$ (red lines), assuming $\alpha= 3.5 \pm
0.5$ and $M_1=(0.82\pm 0.08)\,M_\odot$; they decrease by more than one order
of magnitude from the inner to the outer disk (Fig.\,\ref{fig-density}),
are low at the outer disk rim ($\rho_\mathrm{d0} \sim 10^{-7}-10^{-8}\,
\mathrm{g\,cm^{-3}}$, Fig.\,\ref{fig-mdot}), and show a weak dependency
on mass transfer rate ($\rho_\mathrm{d}\sim \dot{M}_2^{1/4}$).
The changes in slope of the $\rho_\mathrm{d}(\dot{M}_2)$ diagram reflect
whether the first ionization of He occurs in the Rosseland-dominated
region (deep atmospheric layers, $\dot{M}_2 \la 1.6\times 10^{15}\,
\mathrm{g\,s^{-1}}$) or in the Planck-dominated region (upper atmospheric
layers, $1.6\times 10^{15}\,\mathrm{g\,s^{-1}}\la \dot{M}_2 \la 2\times
10^{16}\,\mathrm{g\,s^{-1}}$), and if the temperatures in the vertical
disk structure reach the range of the second ionization of He
($\dot{M}_2 \ga 2\times 10^{16}\,\mathrm{g\,s^{-1}}$).
%
%
Corresponding midplane densities of DIM disk models are plotted as blue
lines in Figs.\,\ref{fig-mdot} and \ref{fig-density}, for $\alpha_c=0.02$
and $\alpha_h=0.1$. The outer regions of a DIM disk are two to three
orders of magnitude denser than those of an MTIM disk accross the
thermally-unstable range of $\dot{M}_2$ values, implying that it is
much harder to satisfy the condition for stream penatration in the DIM
framework than in the MTIM framework.

The midplane density of the gas stream is derived by integrating the
stream density over all widths and heights and conserving mass
\citep{hessman99},
\begin{equation}
  \dot{M}_2 = 2\pi\, H_\mathrm{s} W_\mathrm{s}\,\rho_\mathrm{s0}\,V_\mathrm{s}\, ,
  \label{eq-rhostream}
\end{equation}
where $V_\mathrm{s}$ is the velocity of the gas stream,
\begin{equation}
  V_\mathrm{s}= \frac{2\pi a}{P_\mathrm{orb}}\,v_\mathrm{s}(R,q,a) \, ,
\end{equation}
\noindent and the relative velocity $v_\mathrm{s}(R,q,a)$ is computed
together with the gas stream trajectory by solving the equations of
motion in a coordinate system synchronously rotating with the binary
\citep{ls75}, using a 4th order Runge-Kutta algorithm \citep{numerical}
and conserving the Jacobi integral constant to one part in $10^{6}$.
Combining Eqs.(\ref{eq:hessman1}-\ref{eq-sound}) and (\ref{eq-rhostream})
we obtain,
\begin{equation}
  \rho_\mathrm{s0}\propto \frac{\mu_0 \, \dot{M}_2}
  {v_\mathrm{s}(R,q,a)\, a \,P_\mathrm{orb} \,T_2}
  \,\,\, \mathrm{g\,cm^{-3}} \, .
\end{equation}
We assume $q=0.039\pm 0.002$, $a= (0.29\pm 0.01)\,R_\odot$ (C11),
$P_\mathrm{orb}=1698.5$\,s (SB18) and $\mu_0=4.06$. This relation is
plotted in Figs.~(\ref{fig-mdot}) and (\ref{fig-density}) as black lines
for $T_2= 1000$\,K, 1900\,K and 3700\,K.
%
%
As described by \cite{ls75}, the stretching of fluid elements caused
by the rapid increase of $V_\mathrm{s}$ as the stream leaves L1 leads
initially to a rapid decrease of $\rho_\mathrm{s0}$; this is later more
than offset by the shrinking of both $H_\mathrm{s}$ and $W_\mathrm{s}$, and
$\rho_\mathrm{s0}$ increases again as the stream reaches the inner disk
regions.

We are now ready to test the stream penetration scenario.
Let us first consider the DIM framework. Fig.~\ref{fig-mdot} shows that
$\rho_\mathrm{s0}(R_\mathrm{d}) \ll \rho^+(R_\mathrm{d})$ over the whole
range of $\dot{M}_2$ values where the thermal-viscous instability occurs.
The stream penetration condition $\rho_\mathrm{s0}(R_\mathrm{d}) >
\rho^+(R_\mathrm{d})$ can only be met at extremely large mass transfer
rates $\dot{M}_2 > 10^{18}\,\mathrm{g\,s^{-1}}$, more than two orders of
magnitude larger than both the $\dot{M}_2(\mathrm{quiescence})$ range
inferred in Sect.\,\ref{temperature} and the long-term mass transfer
rate $\dot{M}_2\approx 7.5\times 10^{15}\,\mathrm{g\,s^{-1}}$ predicted
by \cite{Deloye2007}, and far into the hot branch where the YZ\,LMi
accretion disk becomes stable against DIM-driven outbursts -- in
marked contrast with its outbursting nature.
The situation is even worse if the comparison is made between
$\rho_\mathrm{s0}(R_\mathrm{d})$ and $\rho_\mathrm{med}(R_\mathrm{d})$.
Given that $\dot{M}_2(\mathrm{EP1}) \sim 7\,\dot{M}_2(\mathrm{EP2})$,
an additional problem is that allowing for gas stream penetration at the
EP2 epoch forcefully implies a scenario of even stronger gas stream
penetration at EP1 -- in clear disagreement with the C11 observations.
There is no reasonable combination of $T_2$ and $\dot{M}_2$ values that
enables gas stream penetration onto a DIM disk in YZ\,LMi while still
allowing for DIM to explain its outbursts.

Let us switch to the MTIM framework.
The upper panel of Fig.~\ref{fig-mdot} shows that the stream penetration
condition $\rho_\mathrm{s0}(R_\mathrm{d}) > \rho_\mathrm{d0}(R_\mathrm{d})$
is satisfied in the MTIM scenario over a wide range of mass transfer
rates ($\dot{M}_2 \ga 1.6\times 10^{15}\,\mathrm{g\,s^{-1}}$ for $\alpha=4$
and $T_2= 1000\,K$).
Given the constancy of the quiescent brightness of YZ\,LMi over the EP1
and EP2 epochs, we may assume that $\dot{M}_2(\mathrm{quiescence})$ is
the same at both epochs.
The lower panel of Fig.~\ref{fig-mdot} shows the computed penetration
radius for $\alpha= 3.5$. It is possible to have gas stream penetration
at the larger $0.65\,R_\mathrm{L1}$ radius (SB18) while preventing its
occurence at the smaller $0.48\,R_\mathrm{L1}$ radius (C11) over most of
the inferred $\dot{M}_2(\mathrm{quiescence})$ range.
Furthermore, because $\rho_\mathrm{s0} \propto \dot{M}_2$ while
$\rho_\mathrm{d0} \propto \dot{M}_2^{1/4}$, the ratio
$\rho_\mathrm{s0}/\rho_\mathrm{d0}$ increases with increasing $\dot{M}_2$,
indicating that gas stream penetration is easier to trigger and is more
pronounced during bursts of enhanced mass transfer -- in good agreement
with the outburst observations discussed in Sect.~\ref{enhanced}.
Fig.~\ref{fig-density} illustrates how the penetration radius moves inwards
when $\dot{M}_2$ is increased from $3\times 10^{15}\,\mathrm{g\,s^{-1}}$
to $4.8\times 10^{16}\,\mathrm{g\,s^{-1}}$; this corresponds to a $\Delta V
\simeq 2.3$~mag outburst, where the fractional optical flux amplitude
is $\Delta f/f\simeq 8$ and $\dot{M}_2(\mathrm{outburst}) \approx (2 \Delta
f/f)\dot{M}_2(\mathrm{quiescence})\approx 16 \dot{M}_2(\mathrm{quiescence})$.
In this case, $R_\mathrm{p}\leq R_\mathrm{circ}$ and a significant
fraction of the stream mass may be deposited directly at the inner disk
regions giving rise to an inside-out outburst.

Therefore, gas stream penetration onto a high-viscosity MTIM disk provides
a plausible explanation for the presence of significant stream emission
during outbursts, as well as the appearence of enhanced stream emission
when the quiescent YZ\,LMi accretion disk is large, over a reasonable
range of $T_2$ and $\dot{M}_2$ values.

\section{Summary and Discussion}\label{conclusions}

From detailed modeling of the vertical disk structure under the gray
atmosphere approximation, we find that MTIM and DIM disk models of YZ\,LMi
are optically thick at all disk radii for the range of $\alpha$, $\dot{M}$
and $\Sigma$ values of interest, and that tidal dissipations effects are
relevant to the outer disk structure when the disk radius is large and
approaches the tidal truncation radius.

We used the GAIA database to find a revised distance estimate to YZ\,LMi
of $d= 815\pm 138$\,pc, larger than the 460-470\,pc estimate of
\cite{Copperwheat2011}. At this larger distance, the UV-optical fluxes of
a 17000\,K white dwarf at the center of an opaque accretion disk are a
factor $\simeq 6$ lower than observed, implying that the central source
of \cite{Copperwheat2011} is not the white dwarf but an emitting region
significantly larger than the white dwarf itself. Radial brightness
temperature distributions derived from accretion disk maps of \cite{SB2018}
using the updated distance estimate show temperatures ranging from
$\simeq 23000$\,K in the inner disk to $\simeq 5000$\,K in the outer disk
regions that can be reasonably well described by steady-state disk models
in the range $\dot{M}=(1.3-6.3)\times 10^{15}\,\mathrm{g\,s^{-1}}$ at
the 1-$\sigma$ limit. The flattening of the observed distributions in
the outer disk regions is well explained by tidal dissipation on a large,
steady-state opaque quiescent disk.

Observational evidence for enhanced emission along the gas stream trajectory
inwards of the disk rim during outbursts \citep{Copperwheat2011,Szypryt2014}
as well as when its quiescent disk is large \citep{SB2018} suggest
the occurence of gas stream overflow or penetration in the accretion disk
of YZ\,LMi at those epochs. We investigated these two possibilities and
found that it is not possible to explain the enhanced stream emission in
terms of stream overflow because the vertical scaleheight of the gas
stream is significantly lower than that of the outer disk rim, and because
there is no combination of parameters which enables stream overflow onto
a larger disk while preventing it onto a smaller one.
The alternative explanation of stream penetration requires the midplane
density of the gas stream to be larger than that of the disk in its outer
regions. This requirement cannot be met by a low-viscosity disk in the DIM
framework because its midplane density is more than an order of magnitude
larger than that of the gas stream at any radius and over the whole range
of mass transfer rates where the thermal-viscous instability occurs.
On the other hand, mass transfer onto a high-viscosity, low-density MTIM
disk allows for gas stream penetration over a wide range of possible donor
star temperatures and mass transfer rates -- both in outburst and when the
quiescent disk is large -- providing a plausible explanation for the
observed enhanced gas stream emission.

The existence of gas stream penetration in the YZ\,LMi accretion disk
is inconsistent with the DIM scenario and suggests that its outbursts
are otherwise powered by bursts of enhanced mass transfer from its donor
star. This suggestions is strengthen by the detailed outburst observations
of \citet{Copperwheat2011} and \citet{Szypryt2014}, which show the eclipse
of an additional, asymmetric and radially extended light source responsible
for a broad dip centred at phase $\simeq -0.25$, seen on different nights
and distinct outbursts. We argue that these features are evidence of gas
stream penetration caused by enhanced mass transfer -- which we propose
is the ultimate cause of the YZ\,LMi outbursts.
YZ\,LMi seems to be the youngest member of the increasing group of dwarf
novae and outbursting AM CVn stars which challenge the prevailing disk
instability model.

The viscosity parameters inferred from the decline timescale of dwarf
nova outbursts are model dependent. Within the DIM framework the outburst
decline is a thermal event, the inward movement of a cooling wave with
velocity $v_c \simeq 10\,\alpha_h c_s(H/R)$ \citep[e.g.,][]{warner2003};
observed decline timescales of a few days lead to $\alpha_h \approx
0.1-0.3$ \citep[e.g.,][]{cannizzo2001,shl2004,coleman16}.
Within the MTIM framework the outburst decline is a viscous event,
the dumping of the excess mass onto the central object at the viscous
drift velocity $v_\mathrm{R}$; observed decline timescales of a few days
lead to $\alpha \approx 1-3$ \citep[for hydrogen-rich disks,][]{bem83}
and $\alpha \approx 3-4$ (for the hydrogen-deficient disk of YZ\,LMi),
one order of magnitude larger than the values inferred assuming DIM.

If dwarf novae and AM CVn stars outbursts are viscous events
(instead of thermal-viscous instability events), than the
inferred viscosity parameters are larger than unity.
In case viscosity arises from hydrodynamical turbulence, this implies
that either the turbulence is anisotropic (with turbulent eddies of
radial mixing-length $l_R/H_\mathrm{d} \simeq 1-4$) or that the
turbulence is supersonic (with $v_\mathrm{turb}/c_s\simeq 1-4$).
In case viscosity arises from magnetic stresses, this implies
that the magnetic energy density, $B^2/8\pi$, is larger than the
thermal energy density, $\rho c_s^2/2$, by factors $1-4$ --- a regime
which has yet to be addressed in studies of angular momentum exchange
in accretion disks \citep[e.g.,][and references therein]{beckwith11}
or of accretion disk dynamos \citep[e.g.,][]{tp92}.

\begin{acknowledgments}
W. S. acknowledges financial support from CNPq/Brazil.
\end{acknowledgments}

%





\end{document}